\documentclass[twocolumn,showpacs,groupedaddress]{revtex4-1}
\usepackage{graphicx}
\usepackage{natbib}
\usepackage{latexsym}
\begin{document}

\title{Global analysis of data on the spin-orbit coupled
$A^{1}\Sigma_{u}^{+}$ and $b^{3}\Pi_{u}$ states of Cs$_{2}$}
\author{Jianmei Bai, E. H. Ahmed, B. Beser, Y. Guan, S. Kotochigova, and A. M. Lyyra}
\affiliation{Physics Department, Temple University, Philadelphia, PA
19122}
\author{S. Ashman}
\altaffiliation{Present address: Department of Physics and Astronomy,
University of Wisconsin - Stevens Point, Stevens Point, WI 54481}
\affiliation{Department of Physics, Lehigh University, Bethlehem,
Pennsylvania 18015}
\author{C. M. Wolfe}
\altaffiliation{Present address: Army Research Laboratory, RDRL-WMP-A,
Aberdeen Proving Ground, MD 21005-5066}
\affiliation{Department of Physics, Lehigh University, Bethlehem,
Pennsylvania 18015}
\author{J. Huennekens}
\affiliation{Department of Physics, Lehigh University, Bethlehem,
Pennsylvania 18015}
\author{Feng Xie}
\affiliation{Department of Physics and Key Lab of Atomic and Molecular
Nanoscience, Tsinghua University, Beijing 100084, China}
\author{Dan Li}
\altaffiliation{Present address: State Key Laboratory of Precision
Spectroscopy and Department of Physics, East China Normal University,
Shanghai 200062}
\affiliation{Department of Physics and Key Lab of Atomic and Molecular
Nanoscience, Tsinghua University, Beijing 100084, China}
\author{Li Li}
\affiliation{Department of Physics and Key Lab of Atomic and Molecular
Nanoscience, Tsinghua University, Beijing 100084, China}
\author{M. Tamanis and R. Ferber}
\affiliation{Laser Center, Department of Physics, University of Latvia,
19 Rains Blvd., Riga LV-1586, Latvia}
\author{A. Drozdova, E. Pazyuk}
\affiliation{Department of Chemistry, Moscow State University,
GSP-2 Leninskie gory 1/3, Moscow 119992, Russia}
\author{A. V. Stolyarov}
\altaffiliation{Email avstol$@$phys.chem.msu.ru}
\affiliation{Department of Chemistry, Moscow State University,
GSP-2 Leninskie gory 1/3, Moscow 119992, Russia}
\author{J. G. Danzl and H.-C. N\"{a}gerl}
\affiliation{Institut f\"{u}r Experimentalphysik und Zentrum f\"{u}r
Quantenphysik, Universit\"{a}t Innsbruck, Technikerstrasse 25, A-6020
Innsbruck, Austria}
\author{N. Bouloufa}
\affiliation{Laboratoire Aim\'{e} Cotton, CNRS, Universit\'{e}
Paris-Sud, B\^{a}t. 505, 91405 Orsay Cedex, France}
\author{O. Dulieu}
\affiliation{Laboratoire Aim\'{e} Cotton, CNRS, Universit\'{e}
Paris-Sud, B\^{a}t. 505, 91405 Orsay Cedex, France}
\author{C. Amiot}
\altaffiliation{Retired from LAC}
\affiliation{Laboratoire Aim\'{e} Cotton, CNRS, Universit\'{e}
Paris-Sud, B\^{a}t. 505, 91405 Orsay Cedex, France}
\author{H. Salami}
\altaffiliation{Present address: Department of Physics, Faculty of
Sciences (V), Lebanese University, Nabatieh, Lebanon}
\affiliation{Department of Physics and Astronomy, SUNY, Stony Brook,
NY 11794-3800}
\author{T. Bergeman}
\affiliation{Department of Physics and Astronomy,
SUNY, Stony Brook, NY 11794-3800}
\date{\today }

\begin{abstract}
We present experimentally derived potential curves
and spin-orbit interaction functions for the strongly perturbed
$A^{1}\Sigma_{u}^{+}$ and $b^{3}\Pi_{u}$ states of the cesium dimer.
The results are based on data from several sources. Laser-induced
fluorescence Fourier transform spectroscopy (LIF FTS) was used some time ago
in the Laboratoire Aim\'{e} Cotton primarily to study the
$X ^{1}\Sigma_{g}^{+}$ state. More recent work at Tsinghua University
provides information from moderate resolution spectroscopy on the lowest
levels of the $b^{3}\Pi_{0u}^{\pm}$ states as well as additional high resolution
data. From Innsbruck University, we have precision data obtained with
cold Cs$_{2}$ molecules. Recent data from Temple University was obtained
using the optical-optical double resonance polarization spectroscopy
technique, and finally, a group at the University of Latvia has added
additional LIF FTS data. In the Hamiltonian matrix, we have used analytic
potentials (the Expanded Morse Oscillator form) with both finite-difference
(FD) coupled-channels and discrete variable representation (DVR) calculations
of the term values.  Fitted diagonal and off-diagonal spin-orbit functions
are obtained and compared with {\it ab initio} results from Temple and
Moscow State universities.
\end{abstract}
\pacs{33.20.t,33.20.Kf,33.15.Pw,31.50.Df,31.15.aj}
\maketitle

\section{introduction}

The lowest electronically excited states of alkali dimers for long
have been of interest as gateway or "window" states for the
excitation of higher singlet or triplet levels \cite{LiLyyra}. Quite
recently, various low excited states of alkali dimers have also been
used as intermediaries in the production of ultracold molecules, as
for RbCs \cite{Sage}, LiCs \cite{Deiglmayr}, KRb \cite{JinKRb, NiKRb,AiKRb},
NaCs \cite{Bignjp}, Rb$_{2}$ \cite{Lang} and Cs$_{2}$
\cite{Dion, Viteau, Danzl, MMark, JGDFar, JGDanzl}. Although molecules
with electric dipole moments have attracted the most interest,
homonuclear species, such as Cs$_{2}$, also offer the interesting conceptual
challenge of non-spherically symmetric particles in a
condensate \cite{Derevianko}. Furthermore the cold atoms can be prepared
with fewer lasers than required for dual species systems. Cold Cs$_{2}$
molecules may also be useful in experiments designed to be sensitive to
the electron:proton mass ratio, as discussed in \cite{DeMmu}.

Various new possibilities inherent in cold molecules have prompted
efforts to obtain and analyze the
spectra of alkali dimer molecules.  The lowest excited states above
the $^{2}S + ^{2}S$ limit, namely the $A^{1}\Sigma^{+}_{(u)}$
and $b^{3}\Pi_{(u)}$ states (where the ungerade designation
applies only to homonuclear species), would seem to be especially
interesting in this regard.  However, transitions
from the ground state may require frequencies for which the availability
of laser sources is limited.  Furthermore, spin-orbit perturbative
interactions between the $A$ and $b$ state levels complicate the spectra.
As a route to cold ground state molecules from weakly bound molecules that
are formed on a Feshbach resonance (FR), the $A$ and $b$ states of
homonuclear species do not have the special advantage of possible triplet to
singlet transfer which is possible in heteronuclear species \cite{WCS}
which have no {\it gerade/ungerade} selection rules.  In view of these
difficulties, it is noteworthy that, to date, only for Cs$_{2}$ have the $A$
and $b$ states been used as intermediates in the production of cold
molecules \cite{Danzl,MMark,JGDFar,JGDanzl}. The spin-orbit perturbations
are somewhat more severe in species containing Rb or Cs than in the $A$ and
$b$ states of other alkali dimers.
Nevertheless, with Cs$_{2}$, Refs. \cite{Danzl,MMark,JGDFar,JGDanzl} used
the $X^{1}\Sigma_{g}^{+}$ component of a weakly bound molecule formed
on a FR to excite to the $A^{1}\Sigma_{u}^{+}$ component of an upper
state level. Two two-photon steps were used to efficiently connect the
FR molecular state, via $X(v=73, J=2)$ to $X(v=0,J=0)$.

Recently, the methods available for analyzing and modeling spectroscopic
data on highly perturbed states have been extended
\cite{LisK2,FerbNaRb1,MRMK2,TB03,QiNa2,FerbNaRb2,FerbNaCs,KKCs,KKCsb,ODRbCs}
As is evident from the discussion below,
each effort to perform a "global analysis" of all available data
on these two electronic states raises additional questions about what
Hamiltonian elements and functions are required, and  what data is
required to determine the parameters so as to achieve a fit with
residuals comparable to experimental uncertainties. There has been
impresssive progress in data acquisition and analysis for the
heteronuclear $A$ and $b$ states from work on NaK
\cite{AJRNaK1,AJRNaK2,AJRNaK3,SunHNaK,FerNaK} to more recent work on
NaRb \cite{FerbNaRb1,FerbNaRb2}, NaCs \cite{FerbNaCs}, KCs
\cite{KKCs,KKCsb}, and RbCs \cite{TB03,ODRbCs}.  For NaRb, NaCs, and
KCs, vibrational assignments of both states have been reliably
determined, and the perturbative interactions have been modeled to
an accuracy of 0.01 cm$^{-1}$ or better.  Higher-order spin-orbit
effects were included in \cite{FerbNaCs} and in \cite{KKCs}, and
there was quite good agreement between empirically extracted and
{\it ab initio} potentials.  Curiously, for one alkali dimer recently used
for cold molecule work, namely KRb \cite{JinKRb,NiKRb},
spectroscopic data on low levels of the $A^{1}\Sigma^{+}$ and
$b^{3}\Pi$ states appear to be lacking, and optical state transfer
processes leading to cold KRb molecules have used other pathways.

There have also been a succession of studies of the $A$ and $b$ states of the
lighter homonuclear alkali dimer
species Li$_{2}$ \cite{XFLi21,XFLi22,AJRLi2,AMLLi2},
Na$_{2}$ \cite{DemNa2,EffNa2,WhgNa2,KatNa2,Na2opt,QiNa2},
K$_{2}$ \cite{AJRK2,JngK2,JongTh,KimK2,LyyK290,LisK2,MRMK2} and
Rb$_{2}$ \cite{ADVRb2,Sal09}.
For Na$_{2}$ \cite{QiNa2}, data now extend almost continuously from the
lowest vibrational levels to the atomic limit.

One question that arises when considering this summary of spectroscopic data
on the $A$ and $b$ states of Cs$_{2}$ is why hyperfine structure (hfs) is
not reported. Earlier work \cite{DemNa2} on the $A$ and $b$ states of
Na$_{2}$, using a molecular beam, and accessing levels with significant
$^{3}\Pi_{2u}$ character, showed hfs structure spanning several
hundred MHz. The atomic hfs interaction of $^{133}$Cs is larger than
that of $^{23}$Na. Two of the techniques in the present work, namely
optical-optical double resonance polarization spectroscopy, and spectroscopy
with cold ground state Cs$_{2}$ molecules, have very small inherent linewidths.
Nevertheless, the transitions observed here are not expected to exhibit 
hyperfine structure, as we discuss in Section VI below.

This paper presents a detailed model of the energy level
structure of the $A$ and $b$ states of Cs$_{2}$. In section II, we
describe experimental techniques used to obtain high resolution
spectra. Section III describes the model used to fit the data.
Section IV discusses presents {\it ab initio} results for potentials and 
spin-orbit functions, calculated from two different approaches.  Section V 
discusses radiative properties of the $A \sim b$ complex, including certain 
anomalies occurring for mixed singlet-triplet upper states. Section VI 
discusses the possibilities for observing hyperfine structure, while
section VII gives a short summary and conclusion.

\section{The Experimental Data}

This report combines experimental data from several sources.  A substantial part
comes from Fourier transform spectroscopy (FTS) performed in
the Laboratoire Aim\'{e}e Cotton (LAC), Orsay, France, in connection
with the study of the Cs$_{2}$ $X^{1}\Sigma_{g}^{+}$ ground state
as presented in \cite{Verges87,AmiotDulieu}. More recently, FTS data has been
obtained at the University of Latvia in Riga. Additional high resolution data
were obtained in Tsinghua University, Beijing, using fluorescence
excitation spectroscopy. Another valuable component of the data
consists of fluorescence lines from the $2^{3}\Delta_{1g}$ state
\cite{Xie08}, observed at Tsinghua University, which provided
information on the lowest vibrational levels of the $b^{3}\Pi_{0u}$
state, of both parities, $\pm(-1)^{J}$. An ultra-high resolution data
set, limited in absolute accuracy by wavelength meter calibration,
was obtained by excitation of cold Cs$_{2}$ molecules at the University
of Innsbruck, covering a limited range of energies. Finally, because
of a substantial gap in the above data on higher $A/b$ levels, the
Lyyra and Huennekens groups, working at Temple University, employed
optical-optical double resonance (OODR) polarization spectroscopy,
with collisional orientation transfer as in \cite{Sal09},
to partially fill in this gap. Unlike the other data sets, the work
at Temple has not been published previously and therefore the discussion is
more detailed.

\subsection{Data from Fourier transform spectroscopy}

Many of the term values used in the present work are the upper state levels
from transitions observed in LAC and reported in \cite{Verges87} and
\cite{AmiotDulieu}.  Cs atoms in a heat pipe oven at 600K were excited,
in the 1987 study by an argon-ion laser, and in the 2002 work by a
titanium-doped sapphire (Ti:Sa) laser, pumped by an argon-ion laser.  The
fluorescence in the backward direction was collected and sent into a
Fourier spectrometer with a 2m optical path length difference.  Wave
numbers were calibrated by reference to a xenon atomic transition
near 3.5 $\mu$m, with absolute uncertainties between 0.001 and
0.003 cm$^{-1}$. For \cite{AmiotDulieu}, 16,900 transitions were measured
from 113 main fluorescence series and 348 series including rotational
relaxation.  The data are no longer available on the LAC website, but are
posted in the EPAPS file associated with this article \cite{EPAPS}.

At the University of Latvia, FTS data on the $A$ and $b$ states of Cs$_{2}$
have been obtained recently in the process of analyzing the
$A/b \rightarrow X$ laser-induced fluorescence (LIF) spectra of
KCs \cite{KKCs,KKCsb} and RbCs \cite{ODRbCs} molecules. In these spectra,
along with transitions in KCs and RbCs, the $A/b \rightarrow X$ LIF spectra
of Cs$_2$ molecules were present as well. In the experiments, the molecules
were produced in a linear heat-pipe at about 565K temperature and excited by
diode lasers centered at 980nm, 1020nm and 1050nm.
The backward LIF spectra have been recorded by Fourier transform spectrometer
IFS-125HR with 0.03 cm$^{-1}$ resolution. The transition frequencies obtained,
with ca. 0.003 - 0.004 cm$^{-1}$ accuracy, have been added to the $X$ state
term values of Cs$_2$ calculated with the $X$ state potential
of \cite{AmiotDulieu}, corrected by \cite{Coxon}. The estimated overall
uncertainty of term values, accounting for Doppler broadening, is about 0.01 cm$^{-1}$.

The relative LIF intensity distributions have been determined in sufficiently
lengthy progressions, accounting for the spectral sensitivity of the InGaAs
detector as described in Ref. \cite{KKCs}. For comparison with theory, the
intensities were averaged over $P,R$ doublet components and corrected for the
spectral sensitivity of the detector.

\subsection{Fluorescence excitation spectroscopy}

 This approach, employed at Tsinghua University, involved two-photon
excitation to a higher state ($^{3}\Pi_{g}$
or $^{3}\Delta_{g}$) from levels of the $X^{1}\Sigma_{g}^{+}$ state via
levels of the $A^{1}\Sigma_{u}^{+}$ state, which was monitored by observing
decay (direct, or collision-induced) to the $a ^{3}\Sigma_{u}^{+}$ state.
After the double resonance signal was found and optimized, the same
$3^{3}\Pi_{g}(v,J=J'+1)$ [or $2^{3}\Delta_{g}(v,J=J'+1)]$ or
$3^{3}\Pi_{g}(v,J=J'-1)$ [or $2^{3}\Delta_{g}(v,J=J'-1)$] upper level
was excited from $A^{1}\Sigma_{u}^{+}(v',J'+2)$ or
$A^{1}\Sigma_{u}^{+}(v',J'-2)$ intermediate levels. Thus the term values of
the $A^{1}\Sigma_{u}^{+}(v',J'+2)$ or $A^{1}\Sigma_{u}^{+}(v',J'-2)$
levels can be obtained by adding the term values of the
$X^{1}\Sigma_{g}^{+}(v'',J''=J'+1$, or $J'+3$) or
$X^{1}\Sigma_{g}^{+}(v'',J''=J'-1$, or $J'-3$) ground levels
\cite{AmiotDulieu} to the pump
laser frequency with an accuracy of $\sim$0.003 cm$^{-1}$.

\subsection{Spectroscopic observations on cold Cs$_{2}$ molecules}

As part of the effort at the University of Innsbruck to produce ultracold
Cs$_{2}$ molecules in the
rovibronic ground state near quantum degeneracy \cite{Danzl,JGDanzl},
weakly bound molecules were formed by association of atoms on a
Feshbach resonance in a high phase-space density sample. Subsequently,
two STIRAP transitions were used to transfer the population to $X(v=0)$.
A first pair of lasers linked the Feshbach level to the intermediate
level, $X(v=73, J=0$ or $J=2)$,  via an $A/b$ excited state
level near 12,500 cm$^{-1}$. Unfortunately, in the present study there
were insufficient data points near the levels around 12,500 cm$^{-1}$, so
these term values could not be accurately fit in our analysis.

The second two-photon transition starting
from the $X(v=73)$ level involved an $A/b$ intermediate level near 10,000
cm$^{-1}$.  Spectroscopy on the first transition, as described in detail
in Ref. \cite{JGDFar}, was carried out by irradiating the Feshbach
molecules for a given time with light near 1126 nm, stepping the laser
frequency with each cycle of the experiment.  For spectroscopy near 10,000
cm$^{-1}$, molecules were first coherently transferred by STIRAP to
$X(v=73,J=2)$, and then irradiated with light near 1350 nm, as described
in Ref. \cite{MMark}. For detection, the molecules were transferred back
to the original weakly bound state, dissociated, and the resulting atoms
were imaged by absorption imaging.  For STIRAP, the lasers were stabilized
to reach both a short-term linewidth and long-term stability on the kHz level
by locking to optical resonators and additional referencing to a stabilized
optical frequency comb.  However, wavemeter calibration was estimated to
be 0.01 cm$^{-1}$, presenting a limitation on the absolute frequency
accuracy of these transitions.  Ultimately, the radiative linewidth of 
perhaps 1 MHz would limit the resolution.

Approximately 14 $A/b$ levels excited from $X$ state levels near $v$=73 
are included in the data set analyzed in the present work.  These term values
are listed in Table 1 of \cite{MMark} in connection with an earlier
version of the present analysis.

\subsection{Fluorescence from higher-lying states}

These data were obtained at Tsinghua University, and are taken from
Ref. \cite{Xie08}. Two successive diode lasers, each
of 5 MHz linewidth, were used to excite $A ^{1}\Sigma_{u}^{+}
\leftarrow X^{1}\Sigma_{g}^{+}$ and $2^{3}\Delta_{1g} \leftarrow
A ^{1}\Sigma_{u}^{+}$ transitions. Fluorescence from $2^{3}\Delta_{1g}
\rightarrow b^{3}\Pi_{0u}$ was dispersed with a 0.85 m double grating
Spex 1404 monochromator.  The accuracy of the fluorescence lines was
estimated to be about 1.5 cm$^{-1}$, with some better and others worse.

\subsection{Data from OODR polarization spectoscopy, facilitated by
collisional molecular orientation transfer}

V-type optical-optical double-resonance (OODR) polarization
spectroscopy was performed at Temple University to obtain data on
mixed $A/b$ levels in a region of energy not studied previously.
Among different types of polarization
spectroscopy \cite{Teets,Raab,XWang,Nishimiya}, this
technique is well suited to observe selected transitions in the congested
spectra of Cs$_{2}$ since in the single-laser based sub-Doppler
experiments, the assignment of spectral lines is not trivial.
Furthermore, collisional transfer of orientation
produces a series of well-separated spectral lines
associated with nearby levels that have been oriented and polarized.
Another possible approach might be optical triple resonance, as used for
Li$_{2}$ \cite{AMLLi2}, Na$_{2}$ \cite{Na2opt}, and
K$_{2}$ \cite{JngK2,JongTh}. This approach would overcome the unfavorable
Franck-Condon factors mentioned below, but would not yield the
range of $J$ values obtained in the present work, since optimization of
signal to noise through cooling of the sample molecules would probably be
necessary.

\begin{figure}
\includegraphics[scale=0.54]{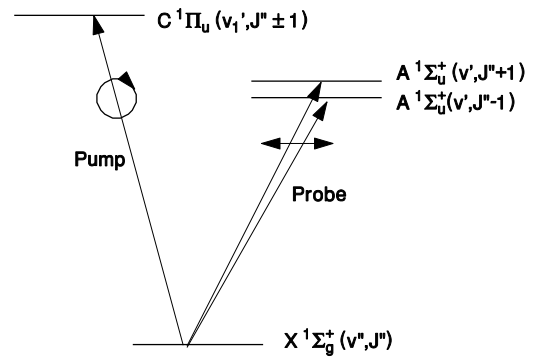}
\caption{A schematic of the transitions used for OODR polarization
spectroscopy. \label{exp1}}
\end{figure}

As in previous implementations of this technique (see \cite{Sal09} and
\cite{QiNa2} and references therein), a circularly
polarized pump laser beam, tuned to a chosen transition, creates an
anisotropic distribution of magnetic sublevel populations (net orientation)
in one or more ground state rovibrational levels.
The counterpropagating linearly polarized probe laser beam can be
considered to be made up of equal parts left and right circular
polarization. When the probe frequency is tuned to a transition sharing
either level with the pump transition, the orientation created by the
pump causes the two circularly polarized probe components to experience
different absorption coefficients (circular dichroism) and different indices of
refraction (circular birefringence).  Consequently, when the probe beam
exits the oven, the two components no longer sum to the initial linear
polarization, but rather the beam has a slight elliptical polarization
and some fraction of it is transmitted through a final crossed
linear polarizer before reaching the detector.  Figure \ref{exp1} shows
schematically the transitions in this technique. More information on the
$C^{1}\Pi_{u}$ state is available from the potentials shown in Fig. \ref{epots}.
A representation of the apparatus is
shown in \cite{QiNa2}.  The experimental setup used here was similar to the
one used previously for Rb$_{2}$ \cite{Sal09}.

Collisional lines help to expand the data field but can only be observed
if orientation is transferred from the ground state level labeled by the
pump laser to a neighboring level.  A quantitative study of the transfer
of population and orientation in collisions of NaK molecules with argon
and potassium atoms has been carried out and reported in \cite{WolfeTh}
and \cite{Wolfeetal}. These references discuss the transfer of orientation
following the discussion of Ref. \cite{Derouard}, and also present a complete
analysis of polarization spectroscopy line shapes, extending the
textbook presentation in \cite{Demtroder}.

In our experiments, the Cs$_{2}$ metal was loaded at the center of a
five-arm heat pipe oven, which was kept at a temperature of 550 K,
with 3 Torr Argon buffer gas. A CR-699-29 tunable single mode laser
with Kiton Red 620 dye was used as the circularly polarized pump
laser, tuned to known \cite{Kasahara}
Cs$_{2}$ $C^{1}\Pi_{u}(v',J' \pm 1) \leftarrow
X^{1}\Sigma_{g}^{+}(v'',J'')$ transitions. A CR-899-29 Ti-Sapphire single
mode laser served as the linearly polarized probe laser. The two
lasers were counter-propagating and overlapped in the center of the
heatpipe with a crossing angle that was minimized so as to maximize
the overlap area.  The power of the pump (probe)
laser was 100mW (20mW) in front of the heatpipe. The spot sizes
(defined as a radius at $1/e^{2}$ intensity) of the pump (1.2 mm)
and probe (0.9mm) laser beams in the interaction region were
measured with a razor blade technique \cite{Skinner}. Two lenses
were placed in the path of each laser beam for collimation inside
the heatpipe.

  A Glan Thompson linear polarizer and a $\lambda$/4 Babinet-Soleil
compensator from Karl Lambrecht Corp. \cite{Lambrecht} were used to make the
pump beam circularly polarized. Two linear polarizers in the probe laser beam
path (one before and one after the heat pipe oven) had relative
transmission axes at a 90 degree angle for
maximum extinction of the probe laser when it was off resonance. A
power meter (Coherent LabMax TO) with a sensitive detector (LM-2)
placed after the second polarizer was used to make fine adjustments
of the orientation of the second polarizer to create an optimal
extinction ratio of 10$^{-6}$ (and then removed for data
acquisition). The observed ratio includes effects of window
birefringence and circular dichroism. A photomultiplier tube
(Hamamatsu R636-10) detected the polarization signal. Its output was
amplified using a lock-in amplifier (Stanford Research Systems SR
850) for phase sensitive detection. The pump laser was modulated at
a frequency of 980 Hz using a mechanical chopper (Stanford Research
Systems SR 540).

The pump laser was calibrated using an iodine atlas \cite{SR,I2atlas},
with estimated accuracy of $\sim$0.004 cm$^{-1}$. The Ti:Sapphire laser
was calibrated by comparing optogalvanic spectroscopy
signals from a uranium lamp to line positions in the uranium atlas
\cite{Uranium} when it scanned above 11000 cm$^{-1}$. The listed
wavenumbers in the Uranium atlas are accurate to $\pm$0.003
cm$^{-1}$. A Burleigh 1600 wavemeter and BOMEM DA8 FTIR were used to
calibrate the Ti:Sapphire frequency when it scanned below 11000
cm$^{-1}$.  The calibrated wavenumbers of the experimental peaks
were added to the term values of their corresponding ground state
levels, calculated from the Dunham parameters of \cite{AmiotDulieu},
to get the $A/b$ state term values.

Because the Franck-Condon factors for transitions of interest were weaker
by a factor of 100 in Cs$_{2}$ than for the Rb$_{2}$ lines in \cite{Sal09},
the signal
to noise ratio was smaller and fewer collisional satellites were observed
than with Rb$_{2}$ ($\Delta J \leq 12$ rather than $\leq 58$).
Also, due to spectral congestion in Cs$_{2}$, the pump transitions were
difficult to isolate. Some of the pump transitions were confirmed by
analyzing the resolved fluorescence recorded by a SPEX model 1404 0.85 m dual
grating monochromator.  Other pump transitions were confirmed by
scanning the pump laser while the probe laser was tuned to a previously
assigned collisional peak.

  26 different $C \leftarrow X$ transitions (ground state
$3 \leq v \leq 8$ and 52 $ \leq J \leq 121$) were confirmed and 20 of them
served as pump transitions in the experiment presented in this work.
These transitions are listed in the EPAPS file \cite{EPAPS}.

\begin{figure}
\includegraphics[scale=0.55]{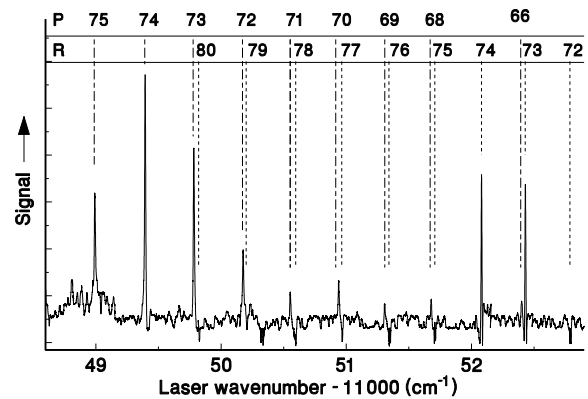}
\caption{OODR polarization spectroscopy signals.  For this scan, the
pump laser was set to 15898.5535 cm$^{-1}$, simultaneously labelling the
ground states of the 
transitions $C^{1}\Pi_{u} (7,73) - X^{1}\Sigma_{g}^{+}$ (6,74) and
$C^{1}\Pi_{u} (7,72) - X^{1}\Sigma_{g}^{+}$ (6,73). In this scan, $P$ branch
lines are always positive going, and $R$ branch lines are negative-going,
except for the strong lines, $R(73)$ and $R(74)$, that reflect off the
baseline as described in the text.  Note that transitions from the
$J$=74 pumped level are relatively strongest. Those associated with
other lower state rotational quantum numbers are due to collisional
transfer of orientation. \label{cs2exp1}}
\end{figure}

 Figure \ref{cs2exp1} shows one OODR polarization spectroscopy data scan.
For homonuclear molecules, the selection rule for collisions is $\Delta J =$
even due to the symmetry properties of the nuclear spin wavefunctions
(for states that are not parity doubled).
However, in the case shown in Fig. 2, the $C^{1}\Pi_{u}$ and
$X^{1}\Sigma_{g}^{+}$ rotational spacings are such that two pump
transitions $C^{1}\Pi_{u}(7,73) - X^{1}\Sigma_{g}^{+}(6,74)$ and
$C^{1}\Pi_{u}(7,72) - X^{1}\Sigma_{g}^{+}(6,73)$ overlap within their
Doppler widths and are therefore simultaneously excited.  Thus Fig. 2
shows two separate collisional series (both characterized by $\Delta J$ =
even) starting from labeled ground state levels (6,74) and (6,73). In this
scan, the off-resonance baseline is artificially created by a non-zero offset
added by the lock-in amplifier.  The $P$ branch lines are positive-going
relative to this offset, and the $R$ branch lines are negative-going here.
The lock-in was used here in the ``R'' mode, in which
the reported signal is the square root of the sum of the squares of the
``in phase'' and ``in quadrature'' components, and hence must always be
positive.  Thus if the $R$ branch lines are particularly strong,
they will reflect off the zero signal baseline, appearing as a central peak
with minima on either side, as shown in Figs. 3b and 3c.  The fact that
$R$ branch and $P$ branch lines have opposite signs (as always occurs
if an offset is used or if only the ``in phase'' channel is recorded) is
helpful for making assignments.  In \cite{Sal09}, the lock-in ``R'' mode
was also used, but without an offset, so $P$ and $R$ lines both appeared
positive going in the polarization scans reported in that work.

\begin{figure}
\includegraphics[scale=0.54]{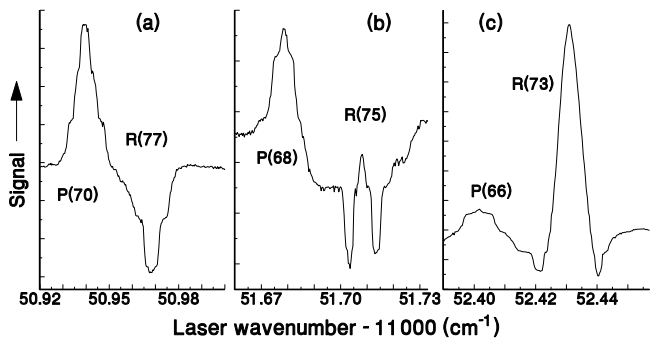}
\caption{OODR polarization spectroscopy line shapes for the particular
conditions used in the present experiments. In each panel,
there is a positive-going $P$ line and an $R$ line. In (a), the $R$
line is weak and entirely negative-going. In  (b) and (c), the $R$
line is stronger and reflects off the zero signal baseline, producing a double
minimum line shape. The vertical scale varies between parts (a)-(c), as
can be seen by comparison with the previous figure. \label{cs2exp2}}
\end{figure}

\subsection{$X$ state term values}

Recently Coxon and Hajigeorgiou \cite{Coxon} have reanalyzed the data of
Amiot and Dulieu \cite{AmiotDulieu} on Cs$_{2}$ $A - X$ transitions to
make a direct fit to a potential of the $X^{1}\Sigma_{g}^{+}$ state,
and hence to obtain improved $X$ state term values. It has come to be
recognized, as asserted in \cite{LeRoyRb2}, that using a direct fit to an
analytic potential is more precise than the semiclassical Dunham procedure.
The differences in this case are less than 0.01 cm$^{-1}$, but nevertheless
significant compared to residuals from least squares fits to $A-b$ state
term values presented below. Supplementary data associated with \cite{Coxon}
give $X$ state term values for transitions in the data set
of \cite{AmiotDulieu}. For other transitions used in this study, we have
calculated $X$ state term values from the potential of \cite{Coxon}.
In doing this, we have used the discrete variable representation (DVR)
described below, to obtain eigenvalues at high $v$ and high $J$ levels without
the use of high-order Dunham coefficients.  When compared with $X$ state
term values listed by \cite{Coxon}, these values agreed to within
4 $\times$ 10$^{-4}$ cm$^{-1}$.

\subsection{Summary of the data}

\begin{figure}
\includegraphics[scale=0.71]{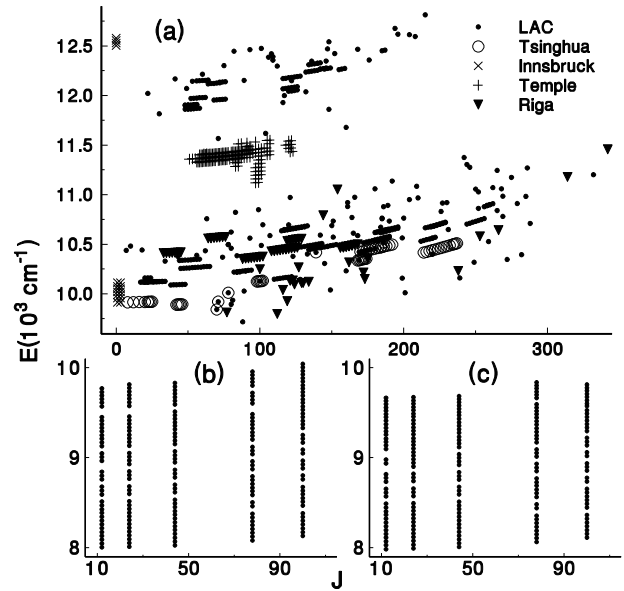}
\caption{Term value data used for this study. (a) Higher resolution
FTS, OODR and cold molecule spectroscopy term values. (b) and (c) display
lower resolution (monochromator) data from Tsinghua University, on $e$ and $f$
parity levels, respectively. Energies in this plot are relative to the
minimum of the $X^{1}\Sigma_{g}^{+}$ ground state. The Innsbruck data above
12,500 cm$^{-1}$ were actually not used in the fit.  However, all other data
shown here fall within the limit $E_{red} = E -0.0091J(J+1) < 12400$ 
cm$^{-1}$, and are included in the least squares fit. Please see Fig. \ref{TAv0}
for a clearer picture of the extent of data near the minimum of the $A$
state, for example. \label{tvalues}}
\end{figure}

Figure \ref{tvalues} summarizes the experimental data used in this
study. It will be noted that the data from Temple University partially fill
in a gap, and that (high-resolution) data from Tsinghua and Riga augment
the data available near the minimum of the $A$ state, as shown more clearly
in Fig. \ref{TAv0}.  The various gaps in the data are associated with regions
of low Franck-Condon factors (see Sec. \ref{tran}) and unavailability of
lasers of suitable wavelength. Data from Innsbruck, at low $J$ values, extend
up to 12,600 cm$^{-1}$ above the minimum of the $X$ state. However, because data
in this region was very limited, we were not able to extend the least
squares fits to the data beyond a reduced energy $E_{red} = E - 0.0091 J(J+1)$
of 12,400 cm$^{-1}$.

When working with data from different sources, there is always a question of
relative calibration errors.  The average residuals (shown in
Table \ref{rmsres} below) indicate that in this case differential
calibration errors are smaller than the rms residuals from the fitting process.
There were few cases of overlapping data to pose a more severe test for
calibration differentials. Six term values measured
by LAC and Tsinghua displayed an average difference of 0.0025 cm$^{-1}$.

\section{Analysis of the Data}

Because the data are relatively sparse in certain regions, the analysis has
required special procedures and furthermore the quality of the fit is not
quite comparable to that of Ref. \cite{KKCs}, for example.  Many  of the
methods are similar to those used in previous studies \cite{Sal09,KKCs}, and
therefore will be summarized briefly, with attention to the special procedures
mandated by the available data set. It is important to note also that for
this relatively heavy alkali dimer, the second-order spin-orbit (SO2)
effects become significant, and will require special attention below.

{\bf Hamiltonian elements.}
The molecular Hamiltonian \cite{Fieldbk}
\begin{eqnarray}
H = H_{BO} +  H_{K} + H_{SO} + H_{rot},
\end{eqnarray}
includes the Born-Oppenheimer potentials $H_{BO}$, radial
kinetic energy $H_{K}$, nuclear rotation $H_{rot}$, and spin-orbit
interaction $H_{SO}$.  Hyperfine structure has not yet been observed, but
in Sec. \ref{hfssec}, we attempt to estimate the magnitude of the
hyperfine structure in the relevant states of Cs$_{2}$.

The observed term values consist of levels of $b^{3}\Pi_{0u}^{\pm}$ below
$v$=0 of the $A ^{1}\Sigma_{u}^{+}$ state, together with many levels
that can be identified as mixtures of $A^{1}\Sigma_{u}^{+}$ levels and $b
^{3}\Pi_{0u}^+$ levels. In the first category are levels of both
parity $(-1)^{J}$ ($e$ symmetry) and parity $-(-1)^{J}$ ($f$ symmetry).
Overall, only a few levels were found to have significant
$^{3}\Pi_{1}$ character and no levels with significant $^{3}\Pi_{2}$ character
were identified. In view of the very limited information on the
$^{3}\Pi_{1}$ levels, our $^{3}\Pi_{0}$ potential contains the
spin-orbit function implicitly. When $^{3}\Pi_{1}$ and $^{3}\Pi_{2}$
levels are considered, the matrix elements of $H_{BO}
+ H_{SO} + H_{rot}+ H_{SO2}$ are \cite{Fieldbk}:
\begin{eqnarray} \label{Ham}
\langle ^{1}\Sigma_{u}^{+}|H|^{1}\Sigma_{u}^{+} \rangle & =
& U_A + (x+2)B + U_{A}^{so2} \nonumber \\
\langle ^{3}\Pi_{0u}^+|H|^{3}\Pi_{0u}^+ \rangle & = & U_{b0}^{+} + (x+2)B
+U_{b0e}^{so2} \nonumber \\
\langle ^{3}\Pi_{0u}^-|H|^{3}\Pi_{0u}^- \rangle & = & U_{b0}^{-} + (x+2)B
+U_{b0f}^{so2} \nonumber \\
\langle ^{3}\Pi_{1u}|H|^{3}\Pi_{1u} \rangle & = & U_{b0}^{+}
+ (x+2)B + \xi^{so}_{b10}  \nonumber \\
\langle ^{3}\Pi_{2u}|H|^{3}\Pi_{2u} \rangle & = & U_{b0}^{+}
+(x-2)B +\xi^{so}_{b10} + \xi^{so}_{b21} \nonumber \\
\langle ^{1}\Sigma_{u}^{+}|H|^{3}\Pi_{0u}^{+} \rangle & =
& - \sqrt{2}\xi^{so}_{Ab0} \\
\langle ^{3}\Pi_{0u}|H|^{3}\Pi_{1u} \rangle & = & - \sqrt{2x} B \nonumber \\
\langle ^{3}\Pi_{1u}|H|^{3}\Pi_{2u} \rangle & = & - \sqrt{2(x-2)} B \nonumber\\
\langle ^{1}\Sigma_{u}^{+}|H|^{3}\Pi_{1u} \rangle & = & - \sqrt{2x} B\eta
\nonumber
\end{eqnarray}
where $x = J(J+1)$, and $H^{T} = H$, where $H^{T}$ is the transpose of $H$.
In the above, the potentials, $U_A$, and $U_{b0}^{\pm}$,
as well as the spin-orbit functions, $\xi^{so}_{Ab0}$ (off-diagonal) and
$\xi^{so}_{b10}$, $\xi^{so}_{b21}$ (on-diagonal); and also
$B \equiv \hbar^{2}/2 \mu r^{2}$ are functions of internuclear distance
$r$ ($\mu$ is the reduced mass). $\xi^{so}_{b10}$ is the $\Omega=1 - 0^{+}$
interval of the $b$ state, and $\xi^{so}_{b21}$ is the $\Omega = 2 - 1$
interval. Both $\xi^{so}_{b10}$ and $\xi^{so}_{b21}$ functions include the
SO2 effect implicitly. The parameter $\eta$ arises from second-order
spin-orbit and electronic-rotational effects, as discussed in \cite{FerbNaCs}.

\begin{figure}
\includegraphics[scale=0.53]{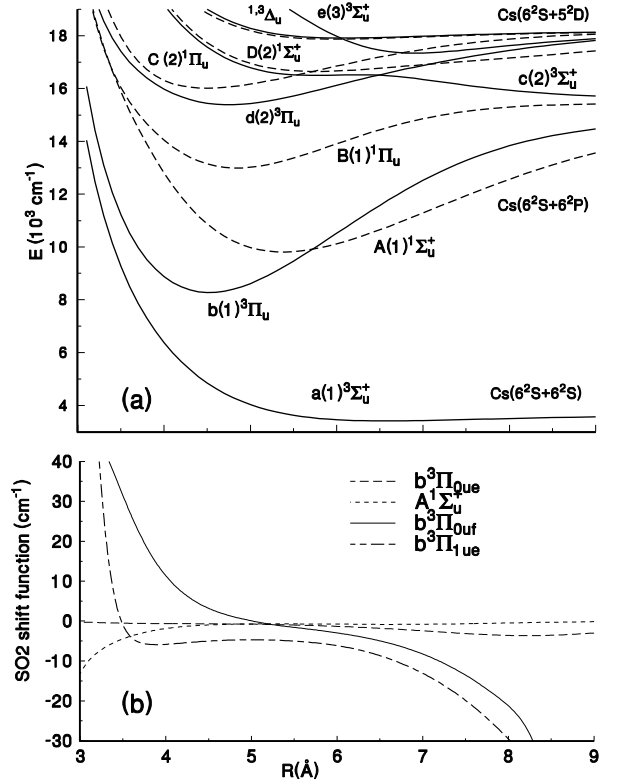}
\caption{(a) Potentials for {\it ungerade} states of Cs$_{2}$
dissociating to the three lowest atomic limits, as calculated from
ECP-CPP-CI methods. (b) Second-order spin-orbit shifts for
$A^{1}\Sigma_{u}^{+}$, $b^{3}\Pi_{0u}^{\pm}$, and $b^{3}\Pi_{1ue}$
calculated from these potential functions and the spin-orbit
coupling functions, using Eqs. \ref{Uaso2}, \ref{Ubeso2}, 
\ref{Ubfso2} and \ref{Ub1eso2}.  \label{epots} }
\end{figure}

The second-order spin-orbit (SO2) terms in the above Hamiltonian elements
have been estimated by {\it ab initio} calculations using the present
quasi-relativistic ECP-CPP-CI procedures described in Sec. \ref{ECPsec}.
These functions are plotted in Fig. \ref{epots}b and given numerically in
the EPAPS \cite{EPAPS} file. The SO2 functions plotted in
Fig. \ref{epots}b are:
\begin{eqnarray} \label{Uaso2}
U_A^{so2}(r)= \frac{2|\xi^{so}_{A-j}|^2}{U_A-U_j};\quad j\in (2)3\Pi_u
\end{eqnarray}
\begin{eqnarray} \label{Ubeso2}
U_{b0e}^{so2}(r)=\frac{2|\xi^{so}_{b0-j}|^2}{U_{b0}^+-U_j};
\quad j\in (2)^1\Sigma^+_u
\end{eqnarray}
\begin{eqnarray} \label{Ubfso2}
U_{b0f}^{so2}(r)=\sum_{j} \frac{2|\xi^{so}_{b0-j}|^2}{U_{b0}^--U_j};
\quad j\in (1-3)^3\Sigma^+_u
\end{eqnarray}
\begin{eqnarray} \label{Ub1eso2}
U_{b1e}^{so2}(r) = \sum_{j} \frac{|\xi^{so}_{b1-j}|^{2}}{U_{
b1}^{+}-U_{j}}; \hspace*{1.3cm} \\ \quad j\in (1-3)^{3}\Sigma^{+}_u;
(1-2)^1\Pi_u \nonumber
\end{eqnarray}
From Fig. \ref{epots}b, it is evident that  $U_{b0f}^{so2}$ is large
and positive for $r < 5.2$ $\AA$, and becomes negative for large $r$.
This behavior can be understood from the
$b^{3}\Pi_{0u}^{-}-a^{3}\Sigma_{u}^{+}$ spin-orbit coupling function,
which shifts the $b^{3}\Pi_{0u}^-$ energies upward at small $r$,
because the $a^{3}\Sigma_{u}^{+}$ potential lies just below the
$b^{3}\Pi_{0u}^{-}$ potential in this region. At larger values of $r$,
the $b^{3}\Pi_{0u}^{-}$ potential approaches various {\it ungerade}
potentials from below, as seen in Fig. \ref{epots}a, and thus the
SO2 shift is negative.

{\bf Analytic potential and spin-orbit forms.} Although an ideal
deperturbation analysis should yield the identical $U_{b0}^{+}$ and
$U_{b0}^{-}$ PECs (potential energy curves) for the triplet state, the
experimental term values
assigned to the $e(+)$ and $f(-)$ parity levels were involved in the
fitting process separately. The empirical PECs of the interacting
$A^1\Sigma^+_u$ and $b^3\Pi_{0u}^{+}$ diabatic states $U_A$, $U_{b0}^+$ were represented analytically
by the Expanded Morse Oscillator (EMO) function \cite{EMO}:
\begin{eqnarray}\label{EMO}
U(r)&=&T_e+{\mathfrak D}_e\left [1 - e^{-\alpha(r-r_e)}\right ]^2; \\
\alpha(r) &=& \sum_{i=0}^N a_i\left (
\frac{r^p-r_{ref}^p}{r^p+r_{ref}^p}\right)^i \nonumber
\end{eqnarray}
converging to the appropriate dissociation limit, namely:
$T_e^A=T_{dis}-{\mathfrak D}_e^A$ for the the singlet $A$-state and
$T_e^{b0}=T_{dis}-\xi_{Cs}^{so}-{\mathfrak D}_e^{b0}$ for the
$\Omega=0$ sub-state of the triplet $b$-state. Here $\xi_{Cs}^{so}
\equiv [E_{6^2{\rm P}_{3/2}}-E_{6^2{\rm P}_{1/2}}]/3$ = 184.6797 cm$^{-1}$
is the empirical spin-orbit constant of the Cs atom in the $6^2P$
state \cite{Steck} while $T_{dis}={\mathfrak D}_e^X+E_{6^2{\rm P}}-
E_{6^2{\rm S}}$ is the energy of the center of gravity of the
Cs$(6^2{\rm P})$ doublet (without hfs splitting) relative to the minimum of
the $X$ state potential. From \cite{Steck} $E_{6^{2}P} - E_{6^{2}S}$ =
11547.6275 cm$^{-1}$. The dissociation energy of the ground state (for the hfs
center-of-gravity) ${\mathfrak D}_e^X=3650.0299(14)$ cm$^{-1}$ from
Ref.\cite{Danzl}, taking into account a correction of -0.0022 cm$^{-1}$
for the energies of $X(v=71-73)$ levels, following \cite{Coxon}.
Thus $T_{dis}$ = 15197.6574 cm$^{-1}$.
Note that because of the $U^{so2}_{A}$ and $U^{so2}_{b0}$ terms, the
effective potential minima differ from the fitted $T_{e}$ values.

Both off-diagonal $\xi^{so}_{Ab0}$ and on-diagonal $\xi^{so}_{b10}$,
$\xi^{so}_{b21}$ empirical spin-orbit coupling functions were approximated
by the Hulburt-Hirschfelder (HH) potential \cite{HHref}
\begin{eqnarray}\label{HHform}
\xi^{so}(r) = \xi_{Cs}^{so}-{\mathfrak D}_e^{so} \left
[2e^{-x}-e^{-2x}[1+cx^3(1+bx)]\right ]; \\
 x = a\left(r/r_e^{so}-1\right) \hspace*{3cm} \nonumber
\end{eqnarray}
determined by the five fitted parameters ${\mathfrak D}_e^{so}$, $r_e^{so}$,
$a$, $b$ and $c$.

The initial EMO and HH parameters of the relevant PECs and SO functions
were estimated using the present quasi-relativistic ECP-CPP-CI electronic
structure calculation. The refined parameters of the PECs and SO coupling
functions $\xi^{so}_{Ab0}$ and $\xi^{so}_{b10}$ were determined iteratively
using a weighted nonlinear least-square fitting (NLSF) procedure. As there
were not data on $b^{3}\Pi_{2u}$, the initial {\it ab initio} estimate
for $\xi^{so}_{b21}$ was used in the least squares fits.

{\bf Fitting procedures.} The spin-orbit coupling function between the
$A^{1}\Sigma_{u}^{+}$ and $b ^{3}\Pi_{0u}^+$ states of Cs$_{2}$ is larger
than the vibrational intervals, thus requiring careful treatment of
perturbative interactions. Initially, the least squares fitting was
performed with a $2\times$2 Hamiltonian matrix, including simply the
$\Omega=0^{+}$ manifolds and the off-diagonal spin-orbit coupling term,
$\xi^{so}_{Ab0}$. The 2 channel fit was crucial in determining nearly
final potentials. The details of the robust fitting procedure used are
given elsewhere \cite{FerbNaCs, KKCs}. Next a 3 $\times$ 3 Hamiltonian
included $^{3}\Pi_{1}$ levels and $\xi^{so}_{b10}$. Final fits were made
with a 4 channel Hamiltonian and $\xi^{so}_{b21}$. Although $^{3}\Pi_{2}$
components were not directly observed, they affected the fitted parameters
through the terms coupling with $^{3}\Pi_{1}$ and thus indirectly to
$^{3}\Pi_{0}^+$.  These multi-channel calculations utilized both
finite-difference (FD) and discrete variable representation (DVR) approaches.
For additional details on the FD and DVR numerical methods, please see
\cite{FerbNaCs} and \cite{Sal09}, respectively. It was determined that
if identical potential and spin-orbit functions are used, the eigenvalues
obtained from the FD and DVR methods are in excellent agreement.

\begin{figure}
\includegraphics[scale=0.57]{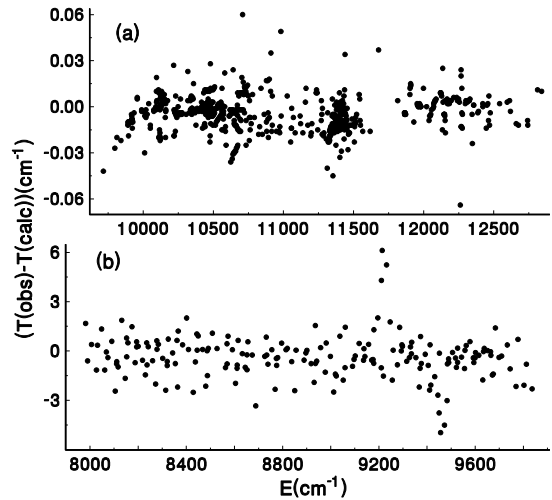}
\caption{Residuals from least squares fits. (a) shows residuals from fits to the high resolution data from all four
sources; (b) shows residuals from the fits to low resolution (monochromator)
data. \label{sigmas}}
\end{figure}

\begin{table}
\caption{Residuals from the least squares fit to Cs$_{2}$ $A/b$ state
data. $N_{p}$ = number of data points. Rms. = Rms residual in cm$^{-1}$;
Ave. = Average residual (observed minus calculated term value), also in
cm$^{-1}$. Tsinghua(LR) denotes the low resolution (monochromator) data
from Tsinghua University (\cite{Xie08}): only $e$ parity levels were
included in this fit. The experimental errors are approximately 1.5
cm$^{-1}$ for Tsinghua(LR) and 0.01 cm$^{-1}$ for the other data sets.
\label{rmsres}}
\begin{tabular}{lrllrll}\hline
Source  &  $N_{p}$ & Rms. & Ave.  \\
\hline LAC \cite{AmiotDulieu, Coxon} & 340 & 0.0154 &  0.0072 \\
LAC \cite{Verges87} &  55 & 0.0198 &  -0.0045 \\
Tsinghua & 58 & 0.0086 &  0.0028 \\
Innsbruck & 14 & 0.0075 & 0.0010 \\
Temple & 161 & 0.0114 & -0.0006  \\
Riga & 75 & 0.0098 & 0.0041 \\
Tsinghua(LR)($e$) & 194 & 1.554 & -0.437 \\
\hline
\end{tabular}
\end{table}

{\bf Fitted parameters, potentials, spin-orbit functions, and term values.}
The $e$ parity term value data was fit with 12(8) potential parameters,
$a_{i}$ in Eq. \ref{EMO}, for the $b^{3}\Pi_{0u}^{+}(A^{1}\Sigma_{u}^{+})$
state, plus $r_{e}$ and $T_{e}$ for each state.  The HH forms
(Eq. \ref{HHform})
used for $\xi^{so}_{Ab0}$ and $\xi^{so}_{b10}$ required 5 fit parameters
each, while the parameters for $\xi^{so}_{b21}$ were provided by {\it
ab initio} calculations (ECP1), The range of the fitted data extended from
7,982 cm$^{-1}$ ($v$=0, $J$=12 of $b^{3}\Pi_{0u+}$) to 12812 cm$^{-1}$,
$J=326$.  But a more meaningful measure is the reduced energy, $E_{res} =
E-0.0091J(J+1)$, which extended only to 12390 cm$^{-1}$.
By contrast, the dissociation limits of the $b^{3}\Pi_{0u}$ and
$A^{1}\Sigma_{u}^{+}$ states lie at 15012.98 cm$^{-1}$ and 15197.66
cm${-1}$, respectively.  The SO2 shift terms, calculated by
Eqs.(\ref{Uaso2}), (\ref{Ubeso2}), (\ref{Ubfso2}) and given numerically in
the EPAPS files \cite{EPAPS}, were added to the $A^{1}\Sigma_{u}^{+}$ and
$b^{3}\Pi_{0u}^{\pm}$ potentials.

Because of the relatively large uncertainty in the
SO2 terms, data for $b^{3}\Pi_{0u}^{-}$ ($f$ parity) were
not included in the fit.

\begin{figure}
\includegraphics[scale=0.54]{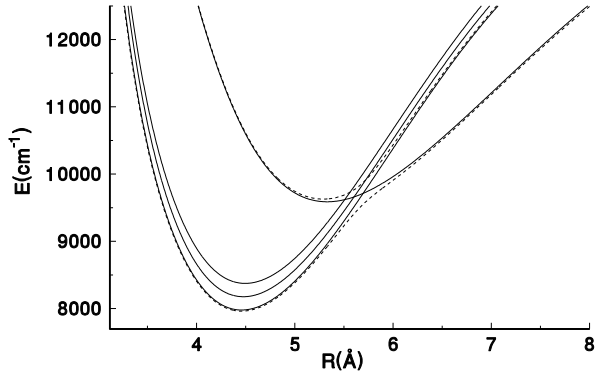}
\caption{Potentials for the Cs$_{2}$ $A^{1}\Sigma_{u}^{+}$ and $b^{3}\Pi_u$
states obtained from fits to the data in this work, over the range sampled
by the data. The $b^{3}\Pi_{2u}$ potential (uppermost of the set of three)
is obtained from {\it ab initio}
calculations only.  The dashed lines indicate the adiabatic $0_{u}^{+}$
potentials, obtained by diagonalizing the diabatic $A$ and $b0$ potentials
with the spin-orbit coupling term, $\xi^{so}_{Ab0}$. \label{pots}}
\end{figure}

The quality of the final fit to the $e$ parity data is indicated by the
plotted residuals in
Fig. \ref{sigmas}, and displayed for the individual data sets in
Table \ref{rmsres}. Table \ref{MLRP} gives the fitted potential parameters
together with results from {\it ab initio} ECP-CPP-CI calculations.
Table \ref{SOF} gives parameters for the empirical off-diagonal and diagonal
spin-orbit functions. Figure \ref{pots} displays the empirical potentials
for the $A^{1}\Sigma_{u}^{+}$ and $b^{3}\Pi_{0u}^+$ states over the range of
the data, namely up to 12,400 cm$^{-1}$ above the minimum of the $X$ state.

Numerical values
for the potentials and spin-orbit functions, parameter listings, calculated
term values and comparisons between calculated and observed term values, and
several plots of calculated and observed term values are all given in the
EPAPS files \cite{EPAPS}.

It should be noted that in fitting the data for $+(e)$ parity levels from
the Tsinghua monochromator (low resolution) data, it was found that
assigning certain terms to $^{3}\Pi_{2u}$ rather than $^{3}\Pi_{0u}^{+}$
reduced the residuals. Ultimately, such assignments were not accepted because
they violate known selection rules for pure (\textbf{a}) Hund's coupling case.
The larger residuals obtained when these levels were assigned to
$^{3}\Pi_{0u}^+$ may be associated with the low intensities of some of the
lines as noted in Ref. \cite{Xie08}, or they may be associated with breakdown
of Hund's case $a$ selection rules due to spin-orbit effects.

\begin{figure}
\includegraphics[scale=0.53]{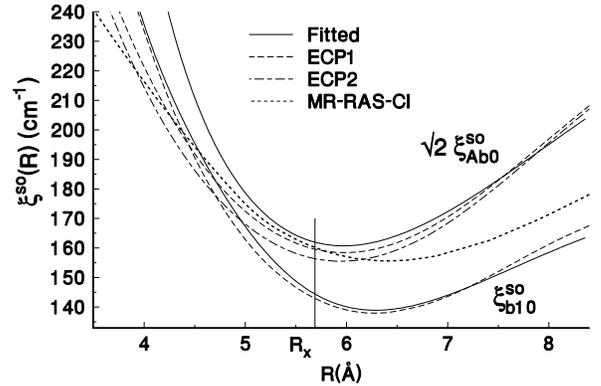}
\caption{Fitted and {\it ab initio} off-diagonal spin-orbit function,
$\sqrt{2}\times \xi^{so}_{Ab0}$, coupling the $A^{1}\Sigma_{u}^{+}$ and
$b^{3}\Pi_{0u}^+$ states, together with the diagonal function,
$\xi^{so}_{b10}$, that gives the fine-structure splitting between
$b^{3}\Pi_{1u}^+$ and $b^{3}\Pi_{0u}^+$. Both MR-RAS-CI and ECP-CPP-CI
procedures were used for the {\it ab initio} functions, as indicated. For
the latter, the upper (lower) line represents results from ECP1 (ECP2) basis
sets. The
vertical line denotes the value $r=r_{x}$ of the potential crossing point.
For $\xi^{so}_{b10}$, the ECP1 and ECP2 results agreed to within the linewidth.
\label{SO}}
\end{figure}

The off-diagonal spin-orbit function $\xi^{so}_{Ab0}$ connecting
$A^{1}\Sigma_{u}^{+}$ and $b^{3}\Pi_{0u}^+$ states is determined
empirically from the perturbation crossings. The fitted parameters are given
in Table \ref{SOF}, and the fitted function is plotted in Fig.\ref{SO},
together with {\it ab initio} functions for which the computational methods
are discussed below. The fitted $\xi^{so}_{Ab0}$ function is primarily
determined by its value at the potential crossing point, denoted $r_{x}$ in
this figure.  It is satisfying that at $r_{x}$, the fitted function is
indeed close to all three {\it ab initio} functions plotted also in this
figure.  For other values of $r$, there are significant differences.

\begin{figure}
\includegraphics[scale=0.58]{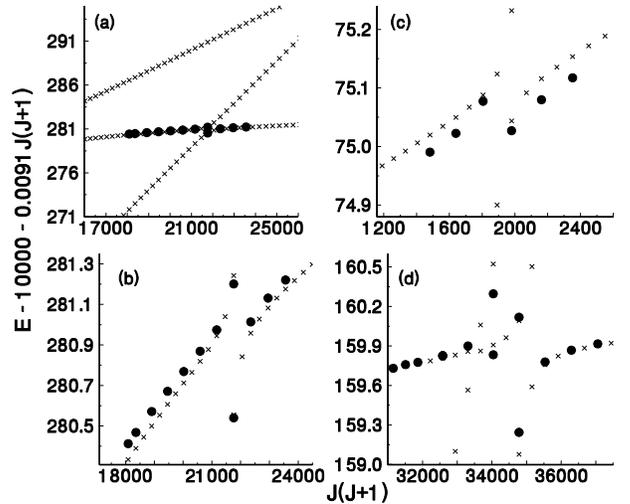}
\caption{Examples of avoided crossings between $\Omega$=1 and
$\Omega=0^{+}$ levels. (a) shows the relatively steeper slope of
$\Omega$=1 levels in a plot of reduced energy vs. $J(J+1)$. (b) is
an enlargement of the crossing region in (a). (c) shows another
crossing region. In (d), a crossing between mixed $\Omega = 0^{+}$ levels
lies very near to an $\Omega=1$ level. Small $\times$s denote calculated
term values, larger filled circles are observed term values in these
plots. \label{fs}}
\end{figure}

The fine structure splitting between the $b^{3}\Pi_{0}$ and $b^{3}\Pi_{1}$
potentials could be deduced from the presently available data only from a
few regions of avoided crossings in the rotational progressions. We have
found four such cases, three of which are shown in Fig. \ref{fs}. From these
deviations in the rotational level energies, we have fit the parameters
(see Table \ref{SOF}) that characterize the fine structure interval. The
fitted $\xi^{so}_{b10}$ function and an {\it ab initio} result are plotted
in Fig. \ref{SO}. The minimum of this empirical fine-structure splitting
function is approximately 139 cm$^{-1}$, as compared with the minima of the
analogous functions for RbCs and KCs, reported to be
71.4 cm$^{-1}$ \cite{ODRbCs} and 78.26 cm$^{-1}$ \cite{KKCs}, respectively,
both based on more complete $b^{3}\Pi_{1}$ data than is presently available
for Cs$_{2}$.  The ECP-CPP-CI {\it ab initio} spin-orbit functions
happen to be in good agreement with the empirical function. It should be
noted that fits of comparable quality were obtained with an empirical
$\xi^{so}_{b10}$ function exhibiting a minimum of $\approx$ 120 cm$^{-1}$,
hence with a different vibrational numbering for the $b^{3}\Pi_{1u}$
manifold. The preference for the function plotted in Fig. \ref{SO} was
based on close agreement with {\it ab initio} results, in view of the good
agreement between empirical and {\it ab initio} SO functions in the case of
KCs \cite{KKCs}.

\begin{table}
\caption{Parameters obtained from a fit to EMO potentials using Eqs.(6)
and (7), together with {\it ab initio} results. In each case, $r_{ref}=5.0$
$\AA$ and $p=3$. $r_{e}$ values are in \AA. $T_{e}$, $D_{e}=T_{dis}-T_e$
and $\omega_{e}$ are in cm$^{-1}$, and the other parameters are dimensionless.
$\eta=-0.08228330$ is defined in Eq.(2). {\it Ab initio} ECP1-CPP-CI values
for $b^{3}\Pi_{0u}$ are obtained from calculated
$U(^{3}\Pi_{1})-\xi^{so}_{b10}$. The $\omega_{e}$ values are obtained from
the second derivative of the potentials at $r=r_{e}$.\label{MLRP}}
\begin{tabular}{rrrrr}\hline \hline
& \multicolumn{2}{c}{Expt.} & \multicolumn{2}{c}{{\it ab initio}} \\
\hline
& $b^{3}\Pi_{0u}^+$ & $A^{1}\Sigma_{u}^{+}$
& $b^{3}\Pi_{0u}^{\pm}$ & $A^{1}\Sigma_{u}^{+}$ \\
\hline
$r_{e}$ & 4.45746 & 5.32913 & \ \ \ 4.5031 & \ \ \ 5.3799  \\
$T_{e}$ & 7977.8479 & 9587.1155 & 8077.8 & 9809.8 \\
$a_{0}$ & 0.522835586 & 0.4332952472 & & \\
$a_{1}$ & 0.135607457 & 0.0431537069 & &  \\
$a_{2}$ & 0.166158407  & -0.01827733744 & & \\
$a_{3}$ & -0.097477983 &   0.12061765970  & & \\
$a_{4}$ & -0.521843942 &  -0.09455975827 & & \\
$a_{5}$ & 1.9047134070 &  0.08598874346 & & \\
$a_{6}$ &  1.737199081 &  0.42659263834 & & \\
$a_{7}$ & -7.992154379  & -0.47446387217 & & \\
$a_{8}$ &  0.076434068   &  & & \\
$a_{9}$ & 8.921002546 &  & & \\
$a_{10}$ & -3.390109463 & & & \\
\hline
$D_{e}$ & 7035.1427 &  5610.5398 & & \\
$\omega_{e}$ & 42.6524 & 33.0055 & 43.133 & 31.953 \\
\hline \hline
\end{tabular}
\end{table}

\begin{table}
\caption{Parameters for the spin-orbit functions in the
Hulburt-Hirschfelder form given in Eq. (8). The parameters for
$\xi^{so}_{Ab0}$ and $\xi^{so}_{b10}$ are from the least squares
fit, while those for $\xi^{so}_{b21}$ are from {\it ab initio}
calculations. In each case,
$\xi^{so}_{Cs}$ = 184.6794 cm$^{-1}$. $D_{e}^{so}$ values are in
cm$^{-1}$, $r_{e}^{so}$ are in \AA, and $a, b$ and $c$ are
dimensionless.\label{SOF}}
\begin{tabular}{rccc}
\hline  & $\xi^{so}_{Ab0}$ & $\xi^{so}_{b10}$ & $\xi^{so}_{b21}$ \\
\hline
$D_{e}^{so}$ & 70.87086 & 46.10116 & 37.2907 \\
$r_{e}^{so}$ & 5.968745 & 6.288671 &  6.341088 \\
$a$ & 2.25268 & 3.34271 & 3.13278 \\
$b$ & 1.221883 & 0.473373 & 0.421783 \\
$c$ & 0.325679 & 0.4050794 & 0.3027172 \\
\hline
\end{tabular}
\end{table}

The last two columns of table \ref{MLRP} give {\it ab initio} ECP1-CPP-CI
results for parameters derived from $U(A^{1}\Sigma_{u}^{+})$ and from
$U(b^{3}\Pi_{0u}^{+})-\xi^{so}_{b10}$.  It is also interesting to compare
results from our least squares fit to the experimental data with earlier
calculations by Spies \cite{Spies}, which were not published but were widely
circulated among people working with Cs$_{2}$. However, since these
calculations were primarily relativistic, they pertain to the adiabatic
potentials, as shown by the dashed lines in Fig. \ref{pots}.  To make a
comparison, we have calculated adiabatic potentials from the experimental
potentials and from the {\it ab initio} ECP1-CPP-CI potentials (by
diagonalizing the potentials plus appropriate spin-orbit functions
as a function of $r$), and then we have extracted values for
$T_{e}, r_{e}$, and $\omega_{e}$, to compare with similar quantities
fit to the potentials of ref. \cite{Spies}.  Table \ref{adia} shows that,
considering the inherent limitations of quantum chemistry calculations,
the earlier {\it ab initio} results were in moderately good agreement
with current experimental results and {\it ab initio results}.

\begin{table}
\caption{Parameters fit to adiabatic $0_{u}^{+}$ potentials from various
sources. Near $r_{e}$, the lower adiabatic potential is close to the
$b^{3}\Pi_{0u}^{+}$ potential, while the upper adiabatic potential is close
to the $A^{1}\Sigma_{u}^{+}$ potential. $T_{e}$ and $\omega_{e}$ are in
cm$^{-1}$, while $r_{e}$ is in \AA. \label{adia}}
\begin{tabular}{rccc}
\hline
& Expt. & ECP1-CPP-CI & Spies  \\
\hline
$T_{e}(lower)$ &  7960.45 & 8064.79 & 8169.57 \\
$r_{e}(lower)$ & 4.4583 & 4.5049 & 4.5028 \\
$\omega_{e}(lower)$ & 42.432 & 42.904 & 43.082 \\
$T_{e}(upper)$ &  9626.64 &  9843.53 & 9598.16 \\
$r_{e}(upper)$ &  5.2903 & 5.3465 & 5.3327 \\
$\omega_{e}(upper)$ & 36.541 & 34.066 & 35.135 \\
\hline
\end{tabular}
\end{table}

\begin{figure}
\includegraphics[scale=0.50]{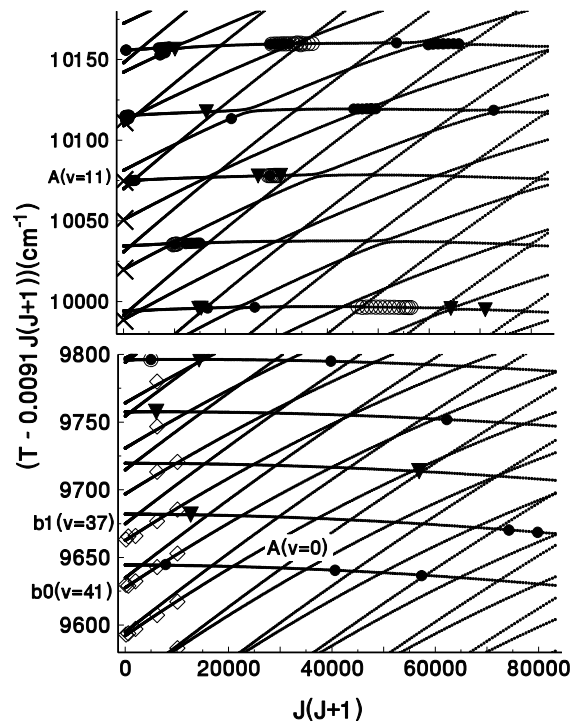}
\caption{Term values near $A(v)$=0 (bottom), showing sparsity of
data, and at higher energies (top), showing somewhat more abundant data
from various sources. Open diamonds indicate monochromator data,
open circles denote high resolution data, also from Tsinghua,
closed circles denote high resolution data from LAC, Xs
denote data from cold molecule spectroscopy from Innsbruck, and
triangles denote data from Riga. The most
steeply sloping eigenvalues are for $\Omega=1$ levels, the least
steeply sloping are for levels that are primarily $A^{1}\Sigma_{u}^{_+}$
in character. \label{TAv0}}
\end{figure}

{\bf Energy level structure - e parity levels.} Figure \ref{TAv0}
displays term values calculated from fitted parameters together with
the input term value data, on a reduced energy scale to flatten the
rotational structure.  One point to note here is that the
experimental data for low levels of the $A$ state are quite sparse
as compared with the study of the $A$ and $b$ states of Rb$_{2}$
\cite{Sal09}, for which it was possible to observe long rotational
progressions that facilitated accurate fitted parameters. The top
part of Fig. \ref{TAv0} displays a region with more ample data. Note
that, despite the strong perturbative coupling, the effective
rotational structure clearly differentiates states that are
primarily $A ^{1}\Sigma_{u}^{+}$, which have the least slope, from
those that are primarily $b^{3}\Pi_{0u}^+$, which have a slope
intermediate between the former and the $b^{3}\Pi_{1u}$ states. In
this regard, please note in the top part of Fig. \ref{TAv0},
that several of the levels observed in Innsbruck University by
excitation from cold molecules in levels of the $X^{1}\Sigma_{g}^{+}$ state,
appear to have primarily triplet character. This feature will be discussed
further in Sec. {\ref{tran}} in connection with transition amplitudes
between mixed $A \sim b$ levels and $X$ state levels.

\begin{figure}
\includegraphics[scale=0.55]{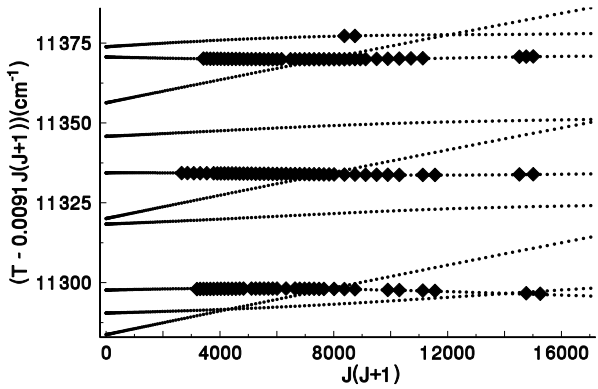}
\caption{Some of the term values (diamonds) measured at Temple University to
partially fill a gap in earlier data. Smaller filled circles denoted calculated
term values. Note that in this energy region, the $0_{u}^{+}$ energy levels
repel each other and do not exhibit narrow avoided crossings, as they do
in the previous figure. \label{TBai}}
\end{figure}

The contrast in rotational structure (slope vs. $J(J+1)$) of $A$ and $b$ state
levels seen in Fig. \ref{TAv0} persists up to approximately $v$=40 of the
$A$ state.  Higher levels, such as those shown in Fig. \ref{TBai} exhibit
a large degree of mixing such that it becomes impossible to assign vibrational
quantum numbers.

\begin{figure}
\includegraphics[scale=0.54]{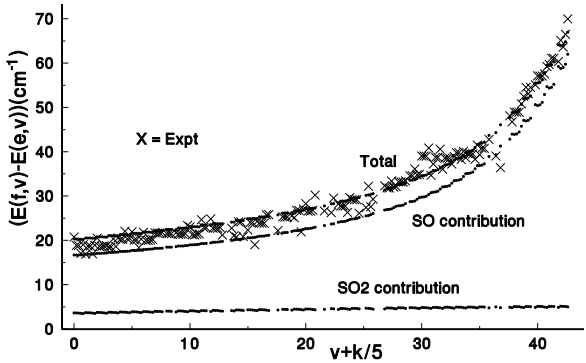}
\caption{Energy difference between $e$ and $f$ parity states for levels
of $b^{3}\Pi_{0u}$, as a function of vibrational level, $v$, and
rotational level. Results are plotted for the 5 values of $J$ observed
in \cite{Xie08}, namely $J$=12, 24, 44, 78 and 100. These $J$ values are
translated into $k$=0,4, and the horizontal axis is $v+k/5$, so as to
spread out the data on different $J$ levels.
The theoretical SO2
contribution is obtained from the second-order spin-orbit coupling functions
and the potential energy differences shown in Fig. \ref{epots}. The
theoretical "SO" contribution is the difference between the calculated
energy with and without $\xi^{so}_{Ab0}$. ``Total'' denotes the sum of
the SO and the SO2 terms. Dots denote theoretical values, $\times$ symbols
denote experimental values.
\label{efdif}}
\end{figure}

{\bf Energy level structure - f parity levels:
Origin of the $e-f$ $\Lambda$-doubling effect
in the b$^3\Pi_{0u}$ sub-state.} As was mentioned above the main contribution
to the $\Omega(\Lambda)$-splitting of the  b$^3\Pi_{0u}$ sub-state \begin{equation}\label{Deltafe}
 \Delta_{fe}=E^f_{vJ}-E^e_{vJ}
\end{equation}
comes from the strong spin-orbit coupling with the nearby singlet $A^1\Sigma_u^+$ state, the "SO" contribution. However, a significant contribution ("SO2")
comes also from the second-order spin-orbit shifts, which are written in
Eq. \ref{Ubfso2}, and plotted in Fig. \ref{epots}b.
Theoretical values for the "SO" contribution are obtained by taking the
difference between eigenvalues calculated with the fitted potentials
with and without the off-diagonal $\xi^{so}_{Ab0}$ coupling function. Calculated
values for the "SO2" contribution are obtained as an expectation value
from the second-order perturbation shift function, $U_{b0f}^{so2}$, shown
in Fig. \ref{epots}b, and the rovibrational wavefunctions calculated from
the $U_{b0}^-$ potential. The results are plotted in Fig. \ref{efdif}.
There is a rapid rise as the $b^{3}\Pi_{0u}$ levels approach $A(v=0)$. The
contribution of SO2 terms increase more slowly between 3.5 and 5 cm$^{-1}$,
as the $b^{3}\Pi_{0u}^-$ functions sample the SO2 shift function shown in
Fig. \ref{epots}b. Because the SO2 terms are likely to be uncertain by
5-10\% (hence possibly as much as 0.5 cm$^{-1}$), we have not attempted an
empirical fit to the $f$ parity data. There may also be an "intrinsic"
$e-f$ energy difference due to spin-spin interaction terms, but
but we have not attempted to estimate such effects.

\section{{\it ab initio} calculations of potentials and spin-orbit functions}
\label{abinit}

In this section we discuss the methods used for {\it ab initio} calculations
at Moscow State University and at Temple University.

\subsection{Quasi-relativistic ECP-CPP-CI calculations (at
Moscow State University)}
\label{ECPsec}

The potential energy curves, transition dipole moments, spin-orbit
and angular coupling matrix elements between the Cs$_2$ electronic states
converging to the lowest three dissociation limits were evaluated in the
basis of the spin-averaged wavefunctions corresponding to pure Hund's
coupling case (\textbf{a}). The quasi-relativistic matrix elements have
been obtained for a wide range of internuclear distance by using effective
core pseudopotential (ECP) \cite{Dolg}. The core-valence correlation has
been taken in account using a large scale multi-reference configuration
interaction (MR-CI) method \cite{MRCI} combined with semi-empirical core
polarization potential (CPP) \cite{Meyer84}. All calculations were performed
by means of the MOLPRO v.2008 program package \cite{MOLPRO2008}.


To include relativistic effects, the inner core shell of the Cs atom has been
replaced by spin-orbit averaged non-empirical small core 9-electrons ECP,
leaving 18 outer-core and valence electrons of the cesium dimer for
explicit correlation treatment. In order to test the ECP basis set
dependence of the present quasi-relativistic calculations, completely
different shape (ECP1) \cite{Zaitsevskii05} and energy (ECP2) \cite{Lim05}
consistent
basis sets available for the Cs atom have been used. The original
spin-averaged Gaussian basis sets from Refs \cite{Zaitsevskii05, Lim05}
were extended by additional diffuse and polarization functions while the
relevant spin-orbit Gaussian sets were directly borrowed from the above
references.

The molecular orbitals (MOs) of Cs$_2$ derived by the self-consistent
field (SCF) method in the $D_{2h}$ point group symmetry were then optimized
by the solution of the state-averaged complete active space SCF (SA-CAS-SCF)
problem for the lowest (1-5)$^{1,3}\Sigma^+_{u,g}$, (1-3)$^{1,3}\Pi_{u,g}$
and (1)$^{1,3}\Delta_{u,g}$ states taken with equal weights \cite{CASSCF}.
The dynamical correlation effects were introduced by the internally contracted
multi-reference configuration interaction method (MR-CI) \cite{MRCI}. The
respective CAS consisted of the 7$\sigma_{u,g}$, 4$\pi_{u,g}$ and
2$\delta_{u,g}$ optimized MOs. MR-CI was applied for only two valence
electrons keeping the rest frozen, i.e. in a full two-valence electron CI
scheme
while the $l$-independent core-polarization potentials (CPPs) with properly
adjusted cutoff radii were employed to take into account the remaining
core-polarization effects implicitly. The relevant spin-orbit Gaussian basis
set coefficients were scaled in order to reproduce the experimental
fine-structure splitting of the lowest excited Cs$(6^{2}P)$ state \cite{Steck}.

To elucidate the impact of the electron correlation effect on the present
results, the CPP-CI energies and wave functions have been repeatedly
evaluated by means of the lower $C_{2v}$ group symmetry. In this case the
respective CAS was restricted by the 14$\sigma$ and 10$\pi$ optimized MOs.
The results obtained in both $D_{2h}$ and $C_{2v}$ representations are found
to be almost identical. Furthermore, the energies and matrix elements
obtained in the framework of the same CPP-CI procedure by using
ECP1 and ECP2 basis sets coincided with each other to within a few percent.

The resulting PECs for all electronic states
of $u$-symmetry converging to the first, second, and third dissociation limits
are depicted in Fig. \ref{epots}a, while some of the relevant diagonal and
off-diagonal spin-orbit coupling functions are compared in Fig. \ref{SO}.

\subsection{Multi-Reference Restricted Active Space Configuration Interaction
(MR-RAS-CI) calculations (Temple University, by author SK)}

At Temple University, we have performed non-relativistic as well as a
relativistic electronic
structure calculation to determine the strength of the spin-orbit coupling
between the $A^1\Sigma^+_u$ and $b^3\Pi_u$ potentials of Cs$_2$.
One electron occupied orbitals are obtained from an atomic Hartree-Fock or
Dirac-Fock calculation, respectively. Virtual, highly-excited orbitals are
Sturm-type functions. The orbitals are labeled $1s, 2s, 2p$ \dots etc in
analogy with the principal quantum number and orbital angular momentum of
the hydrogen atom.  A configuration interaction based on molecular
determinants selected by a multi-reference restricted active space
(MR-RAS-CI) method is used \cite{Roos}. Details on the implementation of this
method in our calculations of heavy diatomic molecules are given in
Ref.~\cite{Kotochigova}. Finally, for the non-relativistic CI calculation
we have evaluated matrix elements of the spin-orbit operator
\begin{eqnarray}
\hat H_{SO} &=& \frac{\alpha^2}{2} \, \sum_{N}\sum_{i} \,
\frac{Z_N}{r_{iN}^3} \, \vec{l}_i\cdot \vec{s}_i \nonumber \\
&-& \frac{\alpha^2}{2} \, \sum_{i \ne j} \,\frac{1}{r_{ij}^3}
[ \vec{r}_{ij} \times \vec{p}_i ]\cdot(\vec{s}_i \,+\, 2\vec{s}_j)\,,
\label{hso}
\end{eqnarray}
where $\alpha$ is the fine structure constant and $\vec s_i$ is the spin of
electron $i$. The first term of Eq.~(\ref{hso}) is an one-electron operator
that describes spin-orbit interactions between one nucleus and one electron.
Here $r_{iN}$ is the separation between the $i$-th electron and nucleus $N$
with charge $Z_N$ and $\vec l_i$ is the electron orbital angular momentum
relative to nucleus $N$. The second term of Eq.~(\ref{hso}) is a two-electron
operator describing the spin-orbit interaction between electrons. Here
$\vec r_{ij}$ is the separation between electrons $i$ and $j$ and
$\vec p_i$ is the momentum of electron $i$.

The closed or filled orbitals up to  the $4d$ shell of Cs form the core
orbitals of the molecular determinants used in the CI for both the
non-relativistic and relativistic calculation. No excitations are allowed
from these shells.  The $5s^2$ and $5p^6$ shells are core-valence orbitals
and are in the active space from which single and double excitations are
allowed. The $6s$ and $6p$ orbitals are also added to the active space and
single, double, and triple occupancy is allowed.  Finally, we use four each
of the $s, p, d,$ and $f$ virtual Sturm orbitals to complete the active space.
Up to double occupancy is allowed for these virtual orbitals.

The off-diagonal spin-orbit matrix elements of the non-relativistic
calculation is shown in Fig.~\ref{SO}. We find that the one-electron
spin-orbit operator of Eq.~(\ref{hso}) provides
$\sim$ 99 \% of the total value of the spin-orbit coupling.

The relativistic potential calculations give an avoided crossing between
$A$ and $b$ potentials. The smallest energy difference $\Delta E$ is evaluated
at r =5.5 \AA\ and equals 326 cm$^{-1}$. According to degenerate perturbtion
theory, half of this value is equal to the off-diagonal $\xi^{s0}_{Ab0}$
spin-orbit function at the same internuclear distance.  Comparison of
$\xi^{so}_{Ab0}$ at $r$ =5.5 \AA\ obtained by non-relativistic
and relativistic calculations shows a good agreement. To be precise, 
$\xi^{so}_{Ab0}$(5.5 \AA) = 163.0, 161.6, 158.2 and 164.4 cm$^{-1}$, from
MR-RAS-CI, ECP1, ECP2 and experiment, respectively.

\section{Transition amplitudes to $X$ state levels} \label{tran}

The distribution of intensities for transitions from or to the $X^{1}\Sigma_{g}^{+}$ state is important in testing model potentials and wavefunctions, for designing data acquisition procedures, and is also important for designing transition sequences to produce cold Cs$_{2}$ molecules from cold Cs atoms or from Cs Feshbach resonance states. Transition intensities can be evaluated by using non-adiabatic wavefunctions of the $A \sim b$ complex calculated by
FD or by DVR methods, and this section will present examples of each.

{\bf Radiative properties of the $A \sim b$ complex.} To test the reliability
of the deperturbation analysis described above, we have evaluated the
$A \sim b \rightarrow X$ transition probabilities, $I$, and radiative lifetimes
$\tau$ of the $A \sim b$ complex according to the relations
\begin{eqnarray}
I_{A\sim b \rightarrow X} \propto \nu^{4}_{A\sim b \rightarrow X}|\langle
\phi_{A} |d_{AX}|v_{X} \rangle |^{2}  \hspace*{8mm} \\
\frac{1}{\tau_{A\sim b}} = \frac{8 \pi^{2}}{2 \hbar c} \sum_{v_{X}}
\nu^{3}_{A\sim b \rightarrow X}|\langle \phi_{A} | d_{AX}| v_{X}
\rangle |^{2} \nonumber \label{lifetime}
\end{eqnarray}
where $\nu_{A\sim b \rightarrow X} = E^{calc}(J') - E_{X}(v_{X};J_{X})$ is
the wavenumber of the rovibronic $A^{1}\Sigma_{u}^{+}\sim b^{3}\Pi_{u}^{+}
\rightarrow X^{1}\Sigma_{g}^{+}$ transition. $d_{AX}(r)$ is the
{\it ab initio} spin-allowed $A^{1}\Sigma_{u}^{+} - X^{1}\Sigma_{g}^{+}$
transition dipole moment calculated in the present work by the ECP-CPP-CI
method. $E^{calc}(J')$ is the energy and $|\phi_{A}\rangle$ is the
non-adiabatic wave-function of the $A$-state component which were obtained
by the FD method from numerical solution of close coupled radial equations
with the present empirical PECs and SO functions. The rovibronic eigenvalues
$E_{X}$ and eigenfunctions of the $X^{1}\Sigma_{g}^{+}$ state were obtained
by solving the single channel radial equation using the empirical potential
from \cite{Coxon}.

\begin{figure}
\includegraphics[scale=0.54]{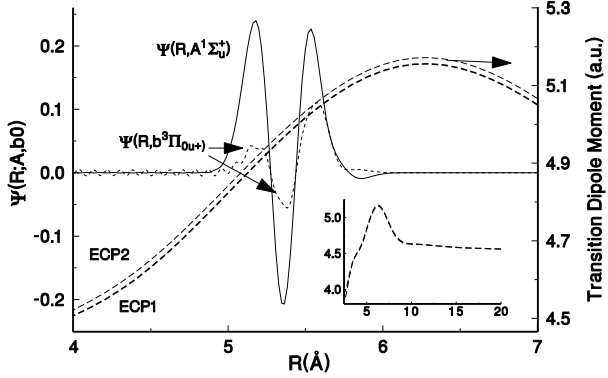}
\caption{(Left axis)Wavefunction components for an excited state level of
$J$=238, which consists mostly of $A^{1}\Sigma_{u}^{+}(v=2)$, but also with
a $b^{3}\Pi_{0u}^+$ component that is shown with dashed lines, indicated by
two arrows. (right axis) Calculated transition dipole
moment between the $A^{1}\Sigma_{u}^{+}$ and $X^{1}\Sigma_{g}^{+}$ state,
in a.u. The ECP1 and ECP2 results from the ECP-CPP-CI method are shown as
discussed in the text. The inset shows theoretical predictions for the
transition dipole moment out to larger values of $r$. \label{wtint}}
\end{figure}

\begin{figure}
\includegraphics[scale=0.54]{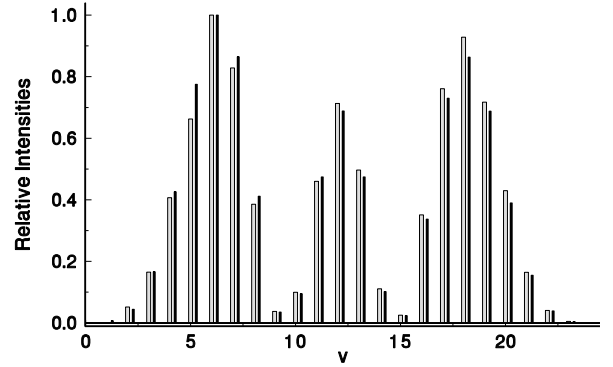}
\caption{Relative intensities for the fluorescence of the
$J^{\prime}=238$ excited state, whose wavefunction components are
shown in the previous figure, to vibrational levels of the
$X^{1}\Sigma_{g}^{+}$ state, as measured in Riga (wider, open bars)
and as calculated by the FD method (narrower, filled bars).
Intensities are normalized to a maximum value of unity in each case.
\label{flint}}
\end{figure}

The Riga group obtained FTS LIF fluorescence intensities over a range of $X$
state vibrational levels from a $J'=$238 level with predominantly
(calculated to be 88\%) $v_{A}=2$ character.
Figure \ref{wtint} shows the two predominant wavefunction components of this
state together with results of two computational results for the
{\it ab initio} transition dipole moment.  The inset shows this moment over
an extended range of $r$. The relative fluorescence intensities in
Figure \ref{flint} show good agreement between observed and calculated
values.

The $\tau_{A\sim b}=26$ $\mu$s value predicted for the metastable ground
$v_b=0$ level of the $b^{3}\Pi_{0u}^+$ state, which has only 0.5\% admixture
of the singlet $A$ state, is very close to the experimental decay rate
$\leq 5 \times 10^{4}$ s$^{-1}$ measured by time-resolved fluorescence
spectra of Cs$_{2}$ dimer molecules immersed in a solid helium
matrix \cite{Moroshkin}. The $\tau_{A\sim b}=0.5$ $\mu$s obtained for the
excited $v_b=37$ level, which has 25\% fraction
of $A^{1}\Sigma_{u}^{+}$ character, also agrees well with the experimental
decay probability of 2.5$\times 10^{3}$ s$^{-1}$ of the $b^{3}\Pi_{u}
\rightarrow X^{1}\Sigma_{g}^{+}$ transition estimated from the kinetics
of Cs$_{2}$ fluorescence measured as a function of temperature and Xe buffer
gas density in \cite{Benedict}.

\begin{figure}
\includegraphics[scale=0.63]{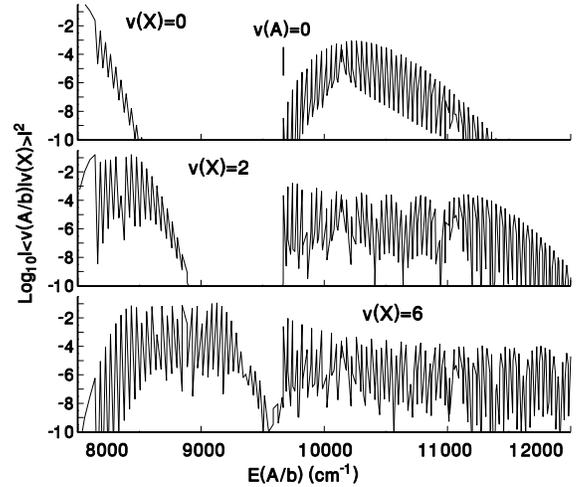}
\caption{Log$_{10}$ of Franck-Condon factors for transitions between 
several levels of the
$X ^{1}\Sigma_{g}^{+}$ state to mixed $A/b$ levels, for $J$=50. Only the
$A ^{1}\Sigma_{u}^{+}$ part of the
mixed wavefunction was used in the calculation. \label{FC}}
\end{figure}

{\bf Transition intensities.} It was noted that the intensities for OODR
polarization spectroscopy experiments at Temple University were much weaker 
than for analogous experiments with Rb$_{2}$.  An explanation for this was
obtained from Franck-Condon (FC) factors,
which showed that over the spectral region of interest  here, the FC
factors were about 100 times smaller than the FC factors that pertained to the
observations of Rb$_{2}$ $A/b \leftarrow X$ transitions reported in
\cite{Sal09}. A plot of FC factors to $A/b$ levels of interest from
$X$ state levels with $v \leq 6$ is shown in Fig. \ref{FC}, for $J$=50. 
Fluctuations in these FC factors are due to the varying fraction of 
$A^{1}\Sigma_{u}^{+}$ character.  Analogous plots for different values
of $J$ are qualitatively similar once the energy scale is adjusted for
the effects of centrifugal distortion.

Note in Fig. \ref{FC} that there are reasonably favorable FC factors
from low levels of the $X^{1}\Sigma_{g}^{+}$ state to low levels of 
$b^{3}\Pi_{0u}^{+}$. For the $X$ state, $r_{e}$ = 4.645 \AA, while for
the $b0$ state, $r_{e}$ = 4.457 \AA \cite{AmiotDulieu}.  Because the $r_{e}$ 
values are similar, the $v$=0 wavefunctions overlap. 
Spin-orbit ($\xi^{so}_{Ab0}$)
mixes in some $A$ state character into the $v$=0 of $b^{3}\Pi_{0u}^{+}$,
producing a significant $(b0-X)(0,0)$ FC factor, as shown in Fig. \ref{FC}.
For higher vibrational levels of $b^{3}\Pi_{0u}^{+}$,
the overlap with $X(v=0)$ is less.  This pattern is repeated for
higher $v$ of the $X$ state, but extending over more $v$ levels of
$b^{3}\Pi_{0u}^{+}$.  The FC factors increase again when the upper state energy
approaches $v$=0 of the $A^{1}\Sigma_{u}^{+}$ state. Thus in Fig. \ref{FC},
there are in general two regions of appreciable FC factors, with a gap
between them.

\begin{figure}
\includegraphics[scale=0.63]{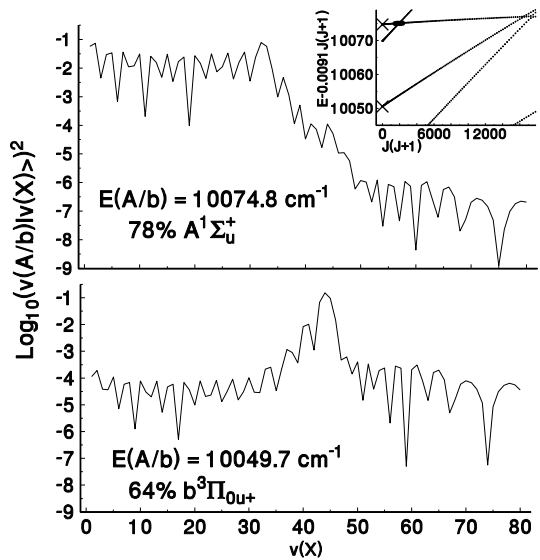}
\caption{Log$_{10}$ of Franck-Condon factors for transitions between two nearby
$A/b$ levels to a range of levels of the $X ^{1}\Sigma_{g}^{+}$
state, showing contrasting behavior. The inset shows that the lower
state has a slope with $J(J+1)$ characteristic of mostly
$b^{3}\Pi_{0u+}$ character, while the level at 10074.8 cm$^{-1}$ has a
slope with $J(J+1)$ characteristic of mostly $A^{1}\Sigma_{u}^{+}$
character. \label{FC2}}
\end{figure}

\begin{figure}
\includegraphics[scale=0.57]{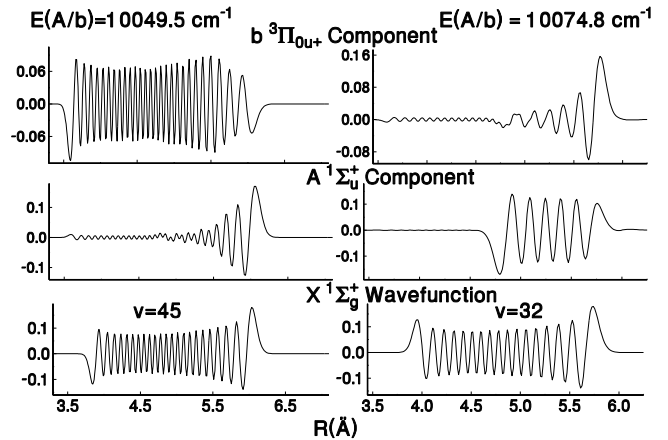}
\caption{Wavefunctions for the excited states in the previous
figure, together with wavefunctions for the $X$ state levels that
give the maximum Franck-Condon overlap with the $A$ state component.
Although the level at 10049.7 cm$^{-1}$ is mostly $b^{3}\Pi_{0u+}$
character, it is intermixed with $A$ state character to give a
substantial Franck-Condon factor for a transition from $X (v=45)$.
\label{FC3}}
\end{figure}

As noted above, one remarkable aspect of the transitions observed
in Innsbruck from $X(v \sim 73)$ to $A/b$ state levels was that the
upper level of the strongest transitions often was primarily triplet in
character, as judged by the calculated rotational structure or the
fractional composition.  This behavior was predicted by {\it ab initio}
calculations by N. Bouloufa and O. Dulieu, as well as by calculations with
fitted potentials and DVR wavefunctions. Figure \ref{FC2} shows the
contrasting behavior in transitions between $X$ state levels and
mixed $A/b$ levels, depending on whether $A$ or $b$ character is
dominant. Figure \ref{FC3} shows that a small admixture of
$A^{1}\Sigma_{u}^{+}$ character, with the correct phase, can produce a
significant Franck-Condon overlap even for a level that is predominantly
$b^{3}\Pi_{0u+}$ in character.  Thus the levels that have predominantly
$b$ state character are preferred as intermediate levels in the 4-photon
STIRAP scheme \cite{MMark, JGDFar, JGDanzl}
because they give a much stronger transition on the first
leg of the second two-photon transition and a more balanced distribution
of transition strengths than levels of predominantly $A$ character.

\section{Can hyperfine structure be observed?} \label{hfssec}

Hyperfine structure (hfs) was not observed in the experiments reported here,
even though $A$ and $b$ states of Na$_{2}$ exhibited hyperfine structure
of several hundred MHz \cite{DemNa2} (in a regime in which the fraction of
$b^{3}\Pi_{2u}$ character was significant), and the 6 $^{2}S$ Fermi contact
splitting in Cs is 9192.63 MHz, as compared with the Na 3 $^{2}S$
hfs splitting of 1771.616 MHz \cite{Arimondo}.  Since the OODR polarization
technique with narrowband lasers is inherently Doppler-free, one might ask why
hfs has not been observed in the present work.
Furthermore, the capability of exciting alkali dimers from ultracold ground
states or from Feshbach resonances, either of which are produced
from cooled atoms, should also lead to possibilities for observing hyperfine
structure with negligible Doppler width, in the lowest excited states of
these species.
With these questions in mind, we present a review of available information and
then make some estimates of possible magnitudes of hyperfine structure.

\subsection{Hamiltonian Matrix Elements}

A detailed theory of hfs in molecular $^{3}\Pi$ states was developed in
\cite{Freed}, and applied to molecular beam observations of the
$a^{3}\Pi$ state of $^{13}$CO \cite{Gammon,Saykally}. \cite{Gammon} translated
the parameters of \cite{Freed} into the more common notation of \cite{FF}.
\cite{Saykally} corrected errors in the Hamiltonian matrix of \cite{Gammon}
and also in the forms adapted from \cite{FF}. An alternative
formalism developed by \cite{BVL}, originally intended for application to
the I$_{2}$ molecule, was applied to the $b^{3}\Pi_{u}$ state of
Na$_{2}$ by \cite{DemNa2}. Later, \cite{LiLi}, from observations of
the $1 ^{3}\Delta_{g}$ state of Na$_{2}$, and \cite{Kato}, from data on
mixed $A^{1}\Sigma_{u}^{+}$ and $b^{3}\Pi_{u}$ states of Na$_{2}$,
found that their observations could be explained almost entirely by the
Fermi-contact parameter alone. Other studies of hfs in alkali dimers,
with one exception noted below, have found the Fermi contact parameter
sufficient.  However, we will present the Hamiltonian matrix elements
for all the magnetic dipole parameters, neglecting possible contributions
from electric quadrupole terms.

In terms of the hyperfine parameters defined in \cite{BC}, (similar
to the parameters of \cite{FF}, but with $b_{F} = b + c/3$),
plus two more parameters, $d$ and $e$, not defined in these references,
the elements of the magnetic dipole hfs operator are
\begin{eqnarray} \label{hfs}
\langle ^{3}\Pi_{0u\pm}|H_{MD}|^{3}\Pi_{1u} \rangle &=&
(y/3) (x/2)^{1/2} (c \mp d -3b_{F}) \nonumber \\
\langle ^{3}\Pi_{1u}|H_{MD}|^{3}\Pi_{1u} \rangle &=& ya
\nonumber \\
\langle ^{3}\Pi_{1u}|H_{MD}|^{3}\Pi_{2u} \rangle &=&
(y/3)[(x-2)/2]^{1/2} (c - 3b_{F}) \nonumber \\
 \langle ^{3}\Pi_{1u}|H_{MD}|^{1}\Sigma_{u}^{+} \rangle
&=& y x^{1/2} e \nonumber \\
\langle ^{3}\Pi_{2u}|H_{MD}|^{3}\Pi_{2u} \rangle &=&
2y(a+b_{F}+2c/3)
\end{eqnarray}
where
\begin{eqnarray} \label{hfsZ}
x & = & J(J+1); \ \ \ y = \frac{[F(F+1)-J(J+1) - I(I+1)]}{2J(J+1)};
\nonumber \\
\vec{F} &=& \vec{J} + \vec{I}.
\end{eqnarray}
$a, b_{F}$ and $ c$ are the nuclear spin-orbital angular
momentum interaction, the Fermi-contact interaction, and the
electron spin-nuclear spin dipolar interaction. $d$ and $e$ are
two additional dipole interaction terms, proportional, respectively,
to $D_{11}$ and $D_{01}$ in \cite{Kato} and \cite{DemNa2}.

Below, following \cite{BC}, we define the parameters used above and,
in the second equality, translate them to the parameters used
by \cite{DemNa2} and \cite{Kato}.
Here $\zeta = g_{S}g_{N} \mu_{B} \mu_{N} (\mu_{0}/4 \pi)$, where
$g_{S}(\mu_{B})$ and $g_{N}(\mu_{N})$ are the electron and nuclear
$g$-factors (Bohr magneton and nuclear magneton), respectively, and
$\mu_{0}$ is the vacuum permeability. Sums below
are over the electrons, $i$, and nuclei, $\alpha$.
\begin{eqnarray}
a &=& \sum_{i,\alpha} (2/g_{S}) \zeta  \langle \Lambda=1 |
1/r_{i,\alpha}^{3} | \Lambda=1 \rangle  = G_{11}/2 \nonumber \\
b_{F} &=& \sum_{i,\alpha} \zeta \left(\frac{8 \pi}{3} \right)  \langle
\Lambda=1 |
\delta({r_{i\alpha}}) | \Lambda=1 \rangle = \left( \frac{K_{11}}{2 \sqrt{6}}
\right). \nonumber \\
c &=& \sum_{i,\alpha} \frac{3}{2} \zeta \langle \Lambda = 1 | (3
\cos^{2} \theta_{i,\alpha}^{2}-1)/r_{i,\alpha}^{3} |\Lambda=1
\rangle = \left( \frac{D_{11}}{8 \sqrt{5}} \right) \nonumber \\
d &=& \sum_{i,\alpha} \frac{3}{2} \zeta \langle \Lambda = -1 |
e^{-2 i \phi_{i,\alpha}} \sin^{2}\theta_{i,\alpha}/r^{3}_{i,\alpha}
| \Lambda =1 \rangle \nonumber \\
e &=& ...  = \left( \frac{D_{10}}{8 \sqrt{15}} \right).
\end{eqnarray}

For $e$, the expression in terms of $\theta_{i,\alpha}$ and $r_{i,\alpha}$,
while straightforward, is too lengthy to give here.

\subsection{Estimate of Splitting from the Fermi Contact and Other Terms}

If we write the Fermi contact term as $H_{FmC} =
b_{F} \vec{I} \cdot \vec{S}$, then from \cite{LiLi} and \cite{Kato},
one obtains $b_{F} \sim A_{hf,atom}/4$, where $A_{hf,atom}$ is the
atomic $^{2}S$ hyperfine parameter. For the
Cs $6^{2}S$ state, $A_{hf,atom}$ = 2298.16 MHz \cite{Arimondo}. Hence the
molecular parameter, $b_{F} \sim$ 574.5 MHz.

As in \cite{DemNa2}, we will designate a mixed $0_{u}^{+}$ state by
\begin{eqnarray}
|0_{u}^{+} \rangle = c_{0}|0\rangle + c_{1}|1\rangle + c_{2}|2\rangle
+c_{\Sigma} |\Sigma \rangle,
\end{eqnarray} \label{hfsa}
where 0, 1, 2 denote $^{3}\Pi_{\Omega,u}$ states with
$\Omega=0,1,2$, and $\Sigma$ denotes the $A^{1}\Sigma_{u}^{+}$
state.

Let us consider a state that is nominally $^{3}\Pi_{0u}$. Considering the
Hamiltonian matrix elements in Eqs. \ref{Ham}, there will be
some degree of $^{3}\Pi_{1}$ admixture, which we can estimate as
\begin{eqnarray}
c_{1} = \frac{\sqrt{2x} B}{E(^{3}\Pi_{1}) - E(^{3}\Pi_{0})} \sim
 \frac{\sqrt{2x} B}{\xi^{so}_{b10}}.
\end{eqnarray}
Considering only terms in $b_{F}$ and letting $\hat{F}=F(F+1)$ etc.,
the hyperfine shifts of this nominal $^{3}\Pi_{0}$ state become
\begin{eqnarray}
\langle 0_{u}^{+} | H_{MD} | 0_{u}^{+} \rangle &=& -\frac{xy B
b_{F}}{\xi^{so}_{b10}}
\sim\ \frac{[\hat{F}-\hat{J}-\hat{I}] B b_{F}}{2\xi^{so}_{b10}} \nonumber \\
&\sim&  20 [\hat{F} - \hat{J} - \hat{I}]\rm{kHz},
\end{eqnarray}
assuming $B \sim 1 \times 10^{-2}$ cm$^{-1}$ and $\xi^{so}_{b10} \sim$
150 cm$^{-1}$. For $J$=0, the result is 0. In general, $F = |I-J| ... (I+J)$.
For given $J$, the maximum value of the expression in brackets is $2JI$,
and thus for $I=7$ (the maximum $I$ value for Cs$_{2}$), the maximum value of
this term is $\sim 280 J$ kHz.  In reality, most $^{3}\Pi_{0u+}$ state
levels of interest will be mixed with $A^{1}\Sigma_{u}^{+}$, so this estimate
should be taken as an upper limit.

Thus, although splitting of the Cs atomic 6 $^{2}S$ state is 9.2 GHz,
hyperfine shifts from the Fermi contact term in Cs$_{2}$
$0_{u}^{+}$ $A/b$ states of low $J$ are less than one MHz,
due to the relatively large fine structure splitting
in Cs$_{2}$. Alternatively, one can say that for $\Omega=0$ and
$\Omega=1$, the electron spin precesses rapidly around the internuclear
axis, thereby reducing terms in $\vec{I} \cdot \vec{S}$.

To our knowledge, only one other hfs term besides the Fermi contact term has
been
convincingly determined in alkali dimer states and that is the $c$ term,
which was found to vary from zero to 9\% $b_{F}$ in various vibrational
levels of the $c^{3}\Sigma^{+}$ state of NaK \cite{KowNaK} (the many fitted
parameters of \cite{DemNa2} must be reviewed in light of the conclusions
of \cite{Kato}).  From Eqs. (\ref{hfs}), the $c$ term would affect the
observed hfs shifts to a fraction comparable to that reported by
\cite{KowNaK}. If $^{3}\Pi_{1u}$ levels of Cs$_{2}$ happen to be observed
in cold molecule spectroscopy, then because the
$\langle ^{3}\Pi_{1u}|H_{MD}|^{3}\Pi_{1u} \rangle$
element is diagonal, the term in $a$ might be detected. If $a$ were also
$\sim$ 5\% of $b_{F}$, shifts could be several MHz.

\section{Conclusions and acknowledgments}

In the interest
of providing a model of the $A$ and $b$ state potentials and energy level
structure for use in the experiments directed to the production of cold
molecules, we have made use of data from several sources, obtained for various
purposes, as noted above.  Because the data are relatively sparse in certain
regions, the analysis has required special procedures.

Questions that warrant further study have been noted above.  For example,
it appeared that some of the $b^{3}\Pi_{0u-}$ ($f$ parity) low vibrational
levels,
from the data of \cite{Xie08}, could better be assigned to $^{3}\Pi_{2u}$.
As laser techniques improve, it would be interesting to attempt to
access these levels with high resolution techniques.  Secondly, the
second-order spin-orbit (SO2) corrections to the $b^{3}\Pi_{1u}$ potential
are quite large, and sensitively dependent on the relative potentials of
perturbing states, as shown in Fig. \ref{epots}.  More extensive data on
$b^{3}\Pi_{1u}$ levels might shed light on these SO2 corrections.  And
finally, as discussed in Sec. \ref{hfssec}, it remains a challenge to
observe hyperfine structure in levels with substantial $b^{3}\Pi_{2u}$
character. This also would require enhanced laser techniques, such as
used recently to study $B^{1}\Pi_{u} \rightarrow X^{1}\Sigma_{g}^{+}$
transitions in Cs$_{2}$ \cite{Nishimiya}.

{\it Acknowledgments.} The work in Temple University was supported by NSF
grant PHY 0855502.  SK acknowledges support from AFOSR and from NSF grant
PHY-1005453.  S.A., C.M. and J. H. were supported by NSF grants
PHY-0652938 and PHY-0968898. The work at Stony Brook was supported by NSF
grants PHY0652459 and PHY0968905. The work in Tsinghua University was
supported by NSFC of China, under grant number 20773072. The Moscow team
thanks the Russian Foundation for Basic Researches by the grant
Nr. 10-03-00195 and MSU Priority Direction 2.3.
M.T. and R.F. are grateful to Ilze Klincare, Olga Nikolayeva and Artis
Kruzins for their help in spectra analysis, as well as appreciate the
support from the ESF 2009/0223/1DP/1.1.1.2.0/09/APIA/VIAA/008 project.



\begin{thebibliography}{}
\bibitem{LiLyyra}See, for example, L. Li and A. M. Lyyra, Spectrochim.
Acta A {\bf 55}, 2147 (1999) on the use of window states to excite higher
triplet states of Li$_{2}$ and Na$_{2}$.
\bibitem{Sage}J. M. Sage, S. Sainis, T. Bergeman and D. DeMille, Phys. Rev.
Lett. {\bf 94}, 203001 (2005).
\bibitem{Deiglmayr}J. Deiglmayr, A. Grochola, M. Repp, K. M\"{o}rtlbauer,
C. Gl\"{u}ck, J. Lange, O. Dulieu, R. Wester and M. Weidem\"{u}ller,
Phys. Rev. Lett. {\bf 101}, 133004 (2008).
\bibitem{JinKRb}S. Ospelkaus, A. Pe'er, K.-K. Ni, J. J. Zirbel, B.
Neyenhuis, S. Kotochigova, P. S. Julienne, J. Ye and D. S. Jin, Nature
Physics, {\bf 4}, 622 (2008).
\bibitem{NiKRb}K.-K. Ni, S. Ospelkaus, M. H. G. de Miranda, A Pe'er,
B. Neyenhuis, J. J. Zirbel, S. Kotochigova, P. S. Julienne, D. S. Jin, and
J. Ye, Science {\bf 322}, 231 (2008).
\bibitem{AiKRb}K. Aikawa, D. Akamatsu, M. Hayashi, K. Oasa, J. Kobayashi,
P. Naidon, T. Kishimoto, M. Ueda and S. Inouye, Phys. Rev. Lett. {\bf 105},
203001 (2010).
\bibitem{Bignjp}C. Haimberger, J. Kleinert, P. Zabawa, A. Wakim and
N. P. Bigelow, New J. Phys. {\bf 11}, 055042 (2009).
\bibitem{Lang}F. Lang, K. Winkler, C. Strauss, R. Grimm, and J. Hecker
Denschlag, Phys. Rev. Lett. {\bf 101}, 133005 (2008).
\bibitem{Dion}C. M. Dion, C. Drag, O. Dulieu, B. LaburtheTolra, F.
Masnou-Seeuws and P. Pillet, Phys. Rev. Lett. {\bf 86}, 2253 (2001).
\bibitem{Viteau}M. Viteau, A. Chotia, M. Allegrini, N. Bouloufa, O.
Dulieu, D. Comparat, and P. Pillet, Science {\bf 321}, 232 (2008).
\bibitem{Danzl}J. G. Danzl, E. Haller, M. Gustavsson, M. J. Mark, R. Hart,
N. Bouloufa, O. Dulieu, H. Ritsch and H.-C. N\"{a}gerl, Science {\bf 321},
1062 (2008).
\bibitem{MMark}M. J. Mark, J. G. Danzl, E. Haller, M. Gustavsson, N. Bouloufa,
O. Dulieu, H. Salami, T. Bergeman, H. Ritsch, R. Hart and H.-C. N\"{a}gerl,
Appl. Phys. B {\bf 95}, 219 (2009).
\bibitem{JGDFar}J. G. Danzl, M. J. Mark. E. Haller, M. Gustavsson, N.
Bouloufa, O. Dulieu, H. Ritsch, R. Hart and H.-C. N\"{a}gerl, Faraday
Discuss. {\bf 142}, 283 (2009).
\bibitem{JGDanzl}J. G. Danzl, M. J. Mark. E. Haller, M. Gustavsson, R. Hart,
J. Aldegunde, J. M. Hutson, H.-C. N\"{a}gerl, Nature Physics {\bf 6},
265 (2010).
\bibitem{Derevianko}A. Derevianko, Phys. Rev. A {\bf 67}, 033607 (2003).
\bibitem{DeMmu}D. DeMille, S. Sainis, J. Sage, T. Bergeman, S. Kotochigova,
and E. Tiesinga, Phys. Rev. Lett. {\bf 100}, 043202 (2008); see also
T. Zelevinsky, S. Kotochigova and J. Ye, Phys. Rev. Lett. {\bf 100}, 043201
(2008).
\bibitem{WCS}W. C. Stwalley, Eur. Phys. J. D {\bf 31}, 221 (2004).
\bibitem{LisK2}C. Lisdat, O. Dulieu, H. Kn\"{o}ckel and E. Tiemann, Eur.
Phys. J. D {\bf 17}, 319 (2001).
\bibitem{FerbNaRb1}M. Tamanis, R. Ferber, A. Zaitsevskii, E. A. Pazyuk,
A. V. Stolyarov, H. Chen, J. Qi, H. Wang and W. C. Stwalley, J. Chem.
Phys. {\bf 117}, 7980 (2002).
\bibitem{MRMK2}M. R. Manaa, A. J. Ross, F. Martin, P. Crozet, A. M. Lyyra,
L. Li, C. Amiot and T. Bergeman, J. Chem. Phys. {\bf 117}, 11208 (2002).
\bibitem{TB03}T. Bergeman, C. E. Fellows, R. F. Gutterres and C. Amiot,
Phys. Rev. A, {\bf 67}, 050501 (2003).
\bibitem{QiNa2}P. Qi, J. Bai, E. Ahmed, A. M. Lyyra, S. Kotochigova, A. J.
Ross, C. Effantin, P. Zalicki, J. Vigu\'{e}, G. Chawla, R. W. Field,
T.-J. Whang, W. C. Stwalley, H. Kn\"{o}ckel, E. Tiemann, J. Shang, L. Li,
and T. Bergeman, J. Chem. Phys. {\bf 127}, 044301 (2007).
\bibitem{FerbNaRb2}O. Docenko, M. Tamanis, R. Ferber, E. A. Pazyuk,
A. Zaitsevskii, A. V. Stolyarov, A. Pashov, H. Kn\"{o}ckel and E. Tiemann,
Phys. Rev. A {\bf 75}, 042503 (2007).
\bibitem{FerbNaCs}J. Zaharova, M. Tamanis, R. Ferber, A. N. Drozdova,
E. A. Pazyuk, and A. V. Stolyarov, Phys. Rev. A {\bf 79}, 012508 (2009).
\bibitem{KKCs}A. Kruzins, I. Klincare, O. Nikolayeva, M. Tamanis, R. Ferber,
E. A. Pazyuk, and A. V. Stolyarov, Phys. Rev. A {\bf 81}, 042509 (2010).
\bibitem{KKCsb}M. Tamanis, I. Klincare, A. Kruzins, O. Nikolayeva, R. Ferber,
E. A. Pazyuk, and A. V. Stolyarov, Phys. Rev. A {\bf 82}, 032506 (2010).
\bibitem{ODRbCs}O. Docenko, M. Tamanis, R. Ferber, T. Bergeman, S. Kotochigova,
A. V. Stolyarov, A. deFaria Nogueira and C. B. Fellows, Phys. Rev. A {\bf 81},
042511 (2010).
\bibitem{AJRNaK1}A.J. Ross, C. Effantin, J. d'Incan and R. F. Barrow,
Mol. Phys. {\bf 56}, 903 (1985).
\bibitem{AJRNaK2}A. J. Ross, C. Effantin, J. d'Incan and R. F. Barrow,
J. Phys. B {\bf 19}, 1449 (1986).
\bibitem{AJRNaK3}A. J. Ross, R. M. Clements, and R. F. Barrow, J. Mol.
Spectrosc. {\bf 127}, 546 (1988).
\bibitem{SunHNaK}H. Sun and J. Huennekens, J. Chem. Phys. {\bf 97}, 4714 (1992).
\bibitem{FerNaK}R. Ferber, E. A. Pazyuk, A. J. Stolyarov, A. Zaitsevskii,
H. Chen, H. Wang and W. C. Stwalley, J. Chem. Phys. {\bf 112}, 5740 (2000).
\bibitem{XFLi21}X. Xie and R. W. Field, Chem. Phys. {\bf 99}, 337 (1985).
\bibitem{XFLi22}X. Xie and R. W. Field, J. Mol. Spectrosc.
{\bf 117}, 228 (1986).
\bibitem{AJRLi2}C. Linton, F. Martin, I. Russier, A. J. Ross, P. Crozet,
S. Churassy and R. Bacis, J. Mol. Spectrosc. {\bf 175}, 340 (1996).
\bibitem{AMLLi2}K. Urbanski, S. Antonova, A. M. Lyyra, A. Yiannopoulou and
W. C. Stwalley, J. Chem. Phys. {\bf 104}, 2813 (1996).
\bibitem{DemNa2}J. B. Atkinson, J. Becker and W. Demtr\"{o}der, Chem. Phys.
Lett. {\bf 87}, 92 (1982).
\bibitem{EffNa2}C. Effantin, O. Babaky, K. Hussein, J. d'Incan and R. F.
Barrow, J. Phys. B {\bf 4077} (1985).
\bibitem{KatNa2}H. Kat\^{o}, M. Otani and M. Baba, J. Chem. Phys. {\bf 89},
653 (1988).
\bibitem{Na2opt}A. M. Lyyra, H. Wang, T.-J. Whang, W. C. Stwalley and L. Li,
Phys. Rev. Lett. {\bf 66}, 2724 (1991).
\bibitem{WhgNa2}T.-J. Whang, W. C. Stwalley, L. Li and A. M. Lyyra, J. Chem.
Phys {\bf 97}, 7211 (1992).
\bibitem{AJRK2}A. J. Ross, P. Crozet, C. Effantin, J. d'Incan and R. F.
Barrow, J. Phys. B {\bf 20}, 6225 (1987).
\bibitem{LyyK290}A. M. Lyyra, W. T. Luh, L. Li, H. Wang and W. C. Stwalley,
J. Chem. Phys. {\bf 92}, 43 (1990).
\bibitem{JongTh}G. Jong, Ph. D. Thesis, University of Iowa, 1991 (unpublished).
\bibitem{JngK2}G. Jong, L. Li, T.-J. Whang, A. M. Lyyra, W. C. Stwalley,
M. Li and J. Coxon, J. Mol. Spectrosc. {\bf 155}, 115 (1992).
\bibitem{KimK2}J. T. Kim, H. Wang, C. C. Tsai, J. T. Bahns, W. C. Stwalley,
G. Jong and A. M. Lyyra, J. Chem. Phys. {\bf 102}, 6646 (1995).
\bibitem{ADVRb2}C. Amiot, O. Dulieu and J. Verg\`{e}s, Phys. Rev.
Lett. {\bf 83}, 2316 (1999).
\bibitem{Sal09}H. Salami, T. Bergeman, B. Beser, J. Bai, E. H. Ahmed,
S. Kotochigova, A. M. Lyyra, J. Huennekens, C. Lisdat, A. V. Stolyarov,
O. Dulieu, P. Crozet and A. J. Ross, Phys. Rev. A {\bf 80}, 022515 (2009).
\bibitem{Verges87}J. Verges and C. Amiot, J. Mol. Spectrosc. {\bf 126},
393 (1987).
\bibitem{AmiotDulieu}C. Amiot and O.Dulieu, J. Chem. Phys. {\bf 117},
5155 (2002).
\bibitem{Xie08}F. Xie, D. Li, L. Tyree, L. Li, V. B. Sovkov, V. S. Ivanov,
S. Magnier and A. M. Lyyra, J. Chem. Phys {\bf 128}, 204313 (2008).
\bibitem{EPAPS}Reference to the EPAPS supplementary file.
\bibitem{Coxon}J. A. Coxon and P. G. Hajigeorgiou, J. Chem. Phys.
{\bf 132}, 094105 (2010).
\bibitem{Teets}R. Teets, R. Feinberg, T. W. H\"{a}nsch and A. L.
Schawlow, Phys. Rev. Lett. {\bf 37}, 683 (1976).
\bibitem{Raab}M. Raab, G. H\"{o}ning, W. Demtr\"{o}der and C. R. Vidal,
J. Chem. Phys. {\bf 76}, 4370 (1982).
\bibitem{XWang}X. Wang, J. Magnes, A. M. Lyyra, A. J. Ross, F. Martin,
P. M.Dove and R. J. Le Roy, J. Chem. Phys. {\bf 117}, 9339 (2002.
\bibitem{Nishimiya}N. Nishimiya, Y. Yasuda, T. Yukiya and M. Suzuki,
J. Mol. Spectrosc. {\bf 255}, 194 (2009).
\bibitem{WolfeTh}C. Wolfe, Ph. D. Thesis, Lehigh University (2010)
(unpublished).
\bibitem{Wolfeetal}C. Wolfe, S. Ashman, B. Beser, E. H. Ahmed, J. Bai,
M. Lyyra and J. Huennekens, in preparation (2010).
\bibitem{Derouard}J. Derouard, Chem. Phys. {\bf 84}, 181 (1984). We thank
D. Pritchard and R. W. Field for alerting us to this reference.
\bibitem{Demtroder}W. Demtr\"{o}der, {\it Laser Spectroscopy, Vol. 2}
Springer, Berlin (2008).
\bibitem{Kasahara}S. Kasahara, Y. Hasui, K. Otsuka, M. Baba, W. Demtr\"{o}der,
and H. Kat\^{o}, J. Chem. Phys. {\bf 106}, 4869 (1997).
\bibitem{Lambrecht}Karl Lambrecht Corporation, Chicago, Il.
\bibitem{Skinner}D. Skinner and R. Whitcher, J. Phys. E {\bf 5}, 237 (1972).
\bibitem{I2atlas}The Aim\'{e} Cotton Iodine Atlas, S. Gerstenkorn and P. Luc, Atlas du Spectre
d'Absorption de la Molecule d'Iode, Editions du CNRS, Paris, 1978,
was recalibrated in 1979 by S. Gerstenkorn and P. Luc, Rev. Phys.
Appl. 14, 791 1979.
\bibitem{SR}H. Salami, A. J. Ross, J. Mol. Spectrosc. {\bf 223}, 157 (2005).
\bibitem{Uranium}B. A. Palmer, R. A. Keller, and R. Engleman, Jr.,
"An atlas of uranium emission intensities in a hollow cathode discharge,"
LASL Rep. LA-8251-MS (Los Alamos Scientific Laboratory, Los Alamos, N. M.,
1980)
\bibitem{LeRoyRb2}J. Y. Seto, R. Le Roy, J. Verg\`{e}s,
C. Amiot, J. Chem. Phys. {\bf 113}, 3067 (2000).
\bibitem{Fieldbk}R. W. Field and H. Lefebvre-Brion. {\it The Spectra and
Dynamics of Diatomic Molecules}, Elsevier, Amsterdam, 2004.
\bibitem{EMO}E. G. Lee, J. Y. Seto, T. Hirao, P. F. Bernath, and R. J. Le Roy,
J. Mol. Spectrosc. {\bf 194}, 197 (1999).
\bibitem{MLRa}R. J. Le Roy, Y. Huang and C. Jary, J. Chem. Phys. {\bf 125},
164310 (2006); R. J. Le Roy and R. D. E. Henderson, Mol. Phys. {\bf 105},
663 (2007).
\bibitem{Steck}D. A. Steck, {\it Cesium D Line Data} (2009),
http://steck.us.alkalidata.
\bibitem{HHref}H. M. Hulburt and J. O. Hirschfelder, J. Chem. Phys. {\bf 9},
61 (1941).
\bibitem{Spies}N. Spies, {\it Ph. D. Thesis}, Fachbereich Chemie,
Universit\"{a}t Kaiserslautern.
\bibitem{Dolg} M.Dolg, \emph{Effective Core Potentials}, published in
Modern Methods and Algorithms of Quantum Chemistry, J. Grotendorst (Ed.),
John von Neumann Institute for Computing, J\"{u}lich, NIC Series, Vol. 1,
pp.479-508, (2000); http://www.fz-juelich.de/nic-series/
\bibitem{MRCI} P. J. Knowles and H.-J. Werner, Theor. Chim. Acta, {\bf 84},
95 (1992).
\bibitem{Meyer84} W.M\"{u}ller, J.Flesch, and W.Meyer, J. Chem.
Phys., {\bf 80}, 3257 (1984).
\bibitem{MOLPRO2008} H. -J. Werner, P. J. Knowles, R. Lindh, F. R. Manby,
M. Schutz, P. Celani, T. Korona, G. Rauhut, R. D. Amos, A. Bernhardsson,
A. Berning, D. L. Cooper, M. J. O. Deegan, A. J. Dobbyn, F. Eckert, C. Hampel,
G. Hetzer, A. W. Lloyd, S. J. McNicholas, W. Meyer, M. E. Mura, A. Nicklass,
P. Palmieri, U. Schumann, H. Stoll, A. J. Stone, R. Tarroni, T. Thosteinsson,
MOLPRO, Version 2008.1, a package of ab initio programs. See www.MOLPRO.net.
\bibitem{Zaitsevskii05} A. Zaitsevskii, E. A. Pazyuk, A. V. Stolyarov,
O. Docenko, I. Klincare, O. Nikolayeva,
M. Auzinsh, M. Tamanis and R. Ferber, Phys. Rev. A, \textbf{71}, 012510 (2005).
\bibitem{Lim05} I. S. Lim, P. Schwerdtfeger, B. Metz, and H. Stoll,
J. Chem. Phys., \textbf{122}, 104103 (2005).
\bibitem{CASSCF} H.-J. Werner and P. J. Knowles, J. Chem. Phys. {\bf 82},
5053 (1985).
\bibitem{Roos}P. A. Malmqvist, A. Rendell, and B. Roos, J. Phys. Chem.
{\bf 94}, 5477 (1990).
\bibitem{Kotochigova}S. Kotochigova and E. Tiesinga, J. Chem. Phys.
{\bf 123}, 174304 (2005).
\bibitem{Moroshkin} P. Moroshkin, A. Hofer, V. Lebedev and A. Weis,
J. Chem. Phys. \textbf{133}, 174510 (2010).
\bibitem{Benedict}R. P. Benedict, D. L. Drummond and L. A. Schlie, J.
Chem. Phys. {\bf 70}, 3155 (1979).
\bibitem{Arimondo}E. Arimondo, M. Inguscio and P. Violino, Rev. Mod. Phys.
{\bf 49}, 31 (1977).
\bibitem{Freed}K. Freed,  J. Chem. Phys. {\bf 43}, 4214 (1966).
\bibitem{Gammon}R. Gammon, R. Stern, M. Lesk, B. Wicke and W. Klemperer,
J. Chem. Phys. {\bf 54}, 2136 (1971).
\bibitem{Saykally}R. Saykally, T. Dixon, T. Anderson, P. Szanto and R.
C. Woods, J. Chem. Phys. {\bf 87}, 6423 (1987).
\bibitem{FF}R. A. Frosch and H. M. Foley, Phys. Rev. {\bf 88}, 1337 (1952).
\bibitem{BVL}M. Broyer, J. Vigu\'{e} and J. C. Lehmann, J. de Physique
{\bf 39}, 591 (1978).
\bibitem{LiLi}L. Li, Q. Zhu  and R. W. Field, J. Mol. Spectrosc. {\bf 134},
50-62 (1989).
\bibitem{Kato}H. Kat\^{o}, M. Otani and M. Baba, J. Chem. Phys. {\bf 91},
5124 (1989).
\bibitem{BC}J. Brown and A. Carrington, {\it Rotational Spectroscopy of
Diatomic Molecules}, Cambridge U. Press, Cambridge, UK, 2003.
\bibitem{KowNaK}P. Kowalczyk, J. Chem. Phys. {\bf 91}, 2779 (1989).
\end{thebibliography}
\end{document}